\documentclass[12pt]{article}
\RequirePackage[OT1]{fontenc}
\RequirePackage{float}
\usepackage{amsmath}
\usepackage{amssymb}
\usepackage{bm}
\usepackage{subcaption}
\usepackage[table,xcdraw]{xcolor}
\usepackage{setspace}
\usepackage[margin=1in]{geometry}
\usepackage[english]{babel}
\usepackage{mathrsfs}
\usepackage{subcaption}
\usepackage{amsthm}
\usepackage[authoryear]{natbib}
\usepackage{lipsum}
\usepackage{multirow}
\usepackage{graphicx}
\newtheorem{theorem}{Theorem}[section]

\newtheorem{assumption}{Assumption}
\newtheorem{remark}{Remark}
\newtheorem{proposition}[theorem]{Proposition}
\newtheorem{algorithm}{Algorithm}
\providecommand{\keywords}[1]{\textbf{\textit{Keywords:}} #1}

\title{A permutation-based Bayesian approach for inverse covariance estimation} 
\author{Xuan Cao\footnote{Corresponding author: Xuan Cao, Department of Mathematical Sciences, University of Cincinnati; email: xuan.cao@uc.edu} \and Shaojun Zhang\footnote{Department of Statistics, University of Florida; email: shaojunzhang@ufl.edu} }
\date{ }

\begin{document}
\allowdisplaybreaks
	\doublespacing
	\noindent
\maketitle
\begin{abstract}
	Covariance estimation and selection for multivariate datasets in a high-dimensional regime is a fundamental problem in modern statistics. Gaussian graphical models are a popular class of models used for this purpose. Current Bayesian methods for inverse covariance matrix estimation under Gaussian graphical models require the underlying graph and hence the ordering of variables to be known. However, in practice, such information on the true underlying model is often unavailable. We therefore propose a novel permutation-based Bayesian approach to tackle the unknown variable ordering issue. In particular, we utilize multiple maximum a posteriori estimates under the DAG-Wishart prior for each permutation, and subsequently construct the final estimate of the inverse covariance matrix. The proposed estimator has smaller variability and yields order-invariant property. We establish posterior convergence rates under mild assumptions and illustrate that our method outperforms existing approaches in estimating the inverse covariance matrices via simulation studies.
\end{abstract}
	\section{Introduction} \label{sec1}
	In modern day statistics, datasets where the number of variables is much larger than the number of samples are more pervasive than they have ever been. Especially in recent years, due to advances in science and technology, data from genomics, finance, environmental and marketing applications are being generated at a rapid pace. One of the major challenges in this setting is to formulate models and develop inferential procedures to understand the complex relationships and multivariate dependencies present in these datasets. In high-dimensional settings, the sample covariance matrix can perform rather poorly. To address the challenge posed by high-dimensionality, several promising methods have been proposed in the literature. In particular, methods inducing sparsity in the Cholesky factor of the inverse covariance matrix $\Omega$ have proven to be very effective in applications. The sparsity patterns in the Cholesky factor of $\Omega$ can be uniquely encoded in terms of appropriate graphs. Hence the corresponding models are often referred to as directed acyclic graph (DAG) models.

	In this paper, we focus on Gaussian DAG models. In particular, suppose we have i.i.d. observations ${\bm Y}_1, {\bm Y}_2, \cdots, {\bm Y}_n$ from a $p$-variate normal distribution with mean vector ${\bf 0}$ and covariance matrix $\Sigma$. Let $\Omega = \Sigma^{-1} = LD^{-1}L^T$ be the modified Cholesky decomposition (MCD) of the inverse covariance matrix, i.e., $L$ is a lower triangular matrix with unit diagonal entries, and $D$ is a diagonal matrix with positive diagonal entries. For a DAG model, this normal distribution is assumed to be Markov with respect to a given directed acyclic graph $ \mathcal{D}$ with vertices $\{1,2, \ldots, p\}$. This is equivalent to saying that $L_{ij}  = 0$ whenever $ \mathcal{D}$ does not have a directed edge from $i$ to $j$ (these concepts are discussed in detail in Section \ref{sec2}). 
	
	There exist several approaches in the literature for estimation of $\Omega$ and the underling graph based on Gaussian DAG models. \citet*{PC:2009} introduce a graph-based technique for estimating sparse covariance and inverse covariance matrices by first inferring the underlying DAG using the PC-algorithm in \citep*{PC:2007}, and estimating the DAG-based covariance matrix and its inverse via the MCD approach. The other type of  methods are based on regularized likelihood/pseudolikelihood. \citet*{Chang:Cholesky:2010} propose a parsimonious approach to estimate high-dimensional covariance matrices via MCD using $\ell_1$ penalization. Similar approach has also been adopted in \citep{Shojaie:Michailidis:2010} via the general weighted lasso to estimate the adjacency matrix of DAGs with given ordering. The authors in \citep*{Vandegeer:2013} show that the $\ell_0$-penalized maximum likelihood estimator of the Cholesky factor converges in Frobenius norm in high dimensions under unknown ordering.
	
		On the Bayesian side, when the underlying graph is known, literature exists that explores the posterior convergence rates for Gaussian concentration graph models, which induce sparsity in the inverse covariance matrix $\Omega$. When the underlying graph is unknown and needs to be selected, comparatively fewer works have tackled with asymptotic properties. Recently, \citep*{BLMR:2016} introduce a flexible 
	and general class of `DAG-Wishart' priors with multiple shape 
	parameters, which are adaptations/generalizations of the 
	Wishart distribution in the DAG context. The authors in \citep*{CKG:2017} establish both strong model selection consistency (in the terminology of \citep*{CKG:2017}) and posterior convergence rates for sparse Gaussian DAG models with DAG-Wishart distributions in a high-dimensional regime. However, the known ordering for these variables is required in order to achieve consistency, which can be problematic in practice, especially when the nature ordering is not available or not per-determined in the format of location or time sequence. 
	
	Recently, \citep*{Kang:2017} adopt an improved MCD approach to tackle the variable order issue in estimating sparse inverse covariance matrix and consider an ensemble estimate under multiple permutations of the variable orders in a frequentist framework, which inspires us to propose a Bayesian ensemble estimate for estimation of $\Omega$ with these DAG-Wishart priors. Specifically, we utilize the multiple MAP (maximum a posteriori) estimates of the Cholesky factor under each permutation, and subsequently construct the final estimate of the inverse covariance matrix. To further encourage the sparsity pattern in the estimate, we also adopt the hard thresholding technique also implemented in \citep*{Kang:2017, CKG:2017} to obtain the final estimate. The proposed estimator has small variability and yields order-invariant property. Under mild assumptions, we establish much better posterior convergence rates under DAG-Wishart distributions.
	
	The rest of the paper is structured as follows. Section \ref{sec2} provides background 
	material from graph theory and Gaussian DAG models. In 
	Section \ref{sec:model} we present our proposed Bayesian approach for precision matrix estimation based on permutation and the posterior convergence rates are provided in 
	Section \ref{sec:posterior}. In 
	Section \ref{sec:simulation} we use simulation experiments to illustrate the proposed method, and demonstrate the benefits of our
	Bayesian approach for inverse covariance matrix estimation
	vis-a-vis existing Bayesian and penalized likelihood approaches. We end our paper with a discussion session in Section \ref{sec:discussion}.
	
		\section{Preliminaries}\label{sec2}                                                                                                            
	\noindent
	In this section, we provide the necessary background material from graph theory, 
	Gaussian DAG models, and DAG-Wishart distributions. 
	
	\subsection{Gaussian DAG models} \label{sec2.1}
	
	\noindent
	Throughout this paper, a directed acyclic graph (DAG) $ \mathcal{D} = (V,E)$ consists 
	of the vertex set $V = \{1,\ldots,p\}$ and an edge set $E$ such that there is no directed 
	path starting and ending at the same vertex. For any given parent ordering, where that all the edges are directed from larger vertices to smaller vertices, denote $pa_i( \mathcal D)$ as the set of parents of $i$ to be the collection of all vertices which are larger than $i$ and share an edge with $i$. 
	Similarly, for any given parent ordering, the set of children of $i$, denoted by $chi_i( \mathcal D)$, is the collection of all vertices which are smaller than $i$ and share an edge with $i$. 
	
	A Gaussian DAG model over a given DAG $ \mathcal{D}$, denoted by 
	$ \mathcal{N}_{ \mathcal{D}}$, consists of all multivariate Gaussian distributions which 
	obey the directed Markov property with respect to a DAG $ \mathcal{D}$. In particular, 
	if $  {y}=(y_1, \ldots, y_p)^T \sim N_p(0,\Sigma)$ and $N_p(0,\Sigma) \in 
	 \mathcal{N}_{ \mathcal{D}}$, then $y_i \perp   {y}_{\{i+1,\ldots,p\}\backslash pa_i( \mathcal D)}|
	{y}_{pa_i( \mathcal D)}$ for each $i$. 
	
	Any positive definite matrix $\Omega$ can be uniquely decomposed as $\Omega = 
	LD^{-1}L^T$, where $L$ is a lower triangular matrix with unit diagonal entries, and $D$ 
	is a diagonal matrix with positive diagonal entries. This decomposition is known as the 
	modified Cholesky decomposition of $\Omega$ (see for example 
	\cite{Pourahmadi:2007}). It is well-known that if $\Omega = LD^{-1}L^T$ is the 
	modified Cholesky decomposition of $\Omega$, then $N_p(0,\Omega^{-1}) \in 
	 \mathcal{N}_{ \mathcal{D}}$ if and only if $L_{ij} = 0$ whenever $i \notin pa_j ( \mathcal D)$. In other 
	words, the structure of the DAG $ \mathcal{D}$ is reflected in the Cholesky factor of 
	the inverse covariance matrix. In light of this, it is often more convenient to 
	reparametrize in terms of the Cholesky parameter of the inverse covariance matrix 
	as follows. 
	
	Given a DAG $ \mathcal{D}$ on $p$ vertices, denote $ \mathcal{L}_{ \mathcal{D}}$ as 
	the set of lower triangular matrices with unit diagonals and $L_{ij} = 0$ if $i \notin 
	pa_j( \mathcal D)$, and let $ \mathcal{D}_+^p$ be the set of strictly positive diagonal matrices in 
	$\mathbb{R}^{p \times p}$. We refer to $\Theta_{ \mathcal{D}} =  \mathcal{D}_+^p 
	\times  \mathcal{L}_{ \mathcal{D}}$ as the Cholesky space corresponding to 
	$ \mathcal{D}$, and $(D,L) \in \Theta_{ \mathcal{D}}$ as the Cholesky parameter 
	corresponding to $ \mathcal{D}$. In fact, the relationship between the DAG and the 
	Cholesky parameter implies that 
	$$
	 \mathcal{N}_{ \mathcal{D}} = \{N_p(0,(L^T)^{-1}DL^{-1}):(D,L) \in 
	\Theta_{ \mathcal{D}}\}. 
	$$
	\subsection{DAG-Wishart distribution} \label{sec2.2}
	
	\noindent
	In this section, we specify the multiple shape parameter DAG-Wishart distributions 
	introduced in \cite{BLMR:2016}. First, we provide required notation on matrix. Given a directed 
	graph $ \mathcal{D} = (V,E)$, with $V = \{1, \ldots, q\}$, and a $q \times q$ matrix $A$, 
	denote the column vectors $A_{ \mathcal D .i}^> = (A_{ij})_{j \in pa_{i}( \mathcal D)}$ and 
	$A_{ \mathcal D.i}^{\ge} = (A_{ii}, (A_{ \mathcal D.i}^>)^T)^T.$ Also, let $A_{ \mathcal 
		D}^{>i} = (A_{kj})_{k,j \in pa_{i}( \mathcal D)}$, $$ A_{ \mathcal D}^{ \ge i} = 
	\left[ \begin{matrix}
	A_{ii} & (A_{ \mathcal D.i}^>)^T \\
	A_{ \mathcal D.i}^> & A_{ \mathcal D}^{>i}
	\end{matrix} \right]. 
	$$
	
	\noindent
	In particular, $A_{ \mathcal D.q}^{\ge} = A_{ \mathcal D}^{ \ge q} = A_{qq}$. 
	
	The DAG-Wishart distributions in \cite{BLMR:2016} corresponding to a DAG 
	$ \mathcal{D}$ are defined on the Cholesky space $\Theta_{ \mathcal{D}}$. Given a 
	positive definite matrix $U$ and a $p$-dimensional vector ${\boldsymbol \alpha} 
	( \mathcal{D})$, the (unnormalized) density of the DAG-Wishart distribution on 
	$\Theta_{ \mathcal{D}}$ is given by 
	\begin{equation} \label{a2}
	\exp\{-\frac12\mbox{tr}((LD^{-1}L^T)U)\} \prod_{i=1}^p 
	D_{ii}^{-\frac{\alpha_i ( \mathcal{D})}2},
	\end{equation} 
	
	\noindent
	for every $(D,L) \in \Theta_{ \mathcal{D}}$. Let $\nu_{i}( \mathcal D) = |pa_{i}
	( \mathcal{D})| = |\{j: j>i, (j,i) \in E({ \mathcal{D}})\}|$. If $\alpha_i( \mathcal D) - 
	\nu_i( \mathcal D) >2$, for all $1 \le i \le q$, the density in (\ref{a2}) can be normalized 
	to a probability density, and the normalizing constant is given by 
	\begin{equation} \label{a1}
	\gamma_{ \mathcal{D}}(U,{\boldsymbol \alpha} ( \mathcal{D})) = \prod_{i=1}^{q}
	\frac{\Gamma(\frac{\alpha_i ( \mathcal{D})}2 - \frac{\nu_{i}( \mathcal D)}2 - 
		1)2^{\frac{\alpha_i ( \mathcal{D})}2 - 1}(\sqrt{\pi})^{\nu_{i}( \mathcal D)} 
		det(U_{ \mathcal D}^{>i})^{\frac{\alpha_i ( \mathcal{D})}2 - \frac{\nu_{i}( \mathcal D)}2 - 
			\frac32}}{det(U_{ \mathcal D}^{\ge i})^{\frac{\alpha_i ( \mathcal{D})}2 - \frac{\nu_{i}
				( \mathcal D)}2 - 1}},
	\end{equation}
	
	\noindent
	In this case, we define the following DAG-Wishart density with $q$ shape parameters 
	$\{\alpha_i ( \mathcal{D})\}_{i=1}^q$ which can be used for differential shrinkage 
	of the variables in high-dimensional settings,
	$\pi_{U,   \alpha( \mathcal D)}^{\Theta_{ \mathcal{D}}}$ on the Cholesky space 
	$\Theta_{ \mathcal{D}}$ by 
	\begin{align} \label{a3}
	\pi_{U,   \alpha( \mathcal D)}^{\Theta_{ \mathcal{D}}} (D,L) = 
	\frac{1}{\gamma_{ \mathcal{D}}(U,{\boldsymbol \alpha} ( \mathcal{D}))} \exp\{-\frac12\mbox{tr}
	((LD^{-1}L^T)U)\} \prod_{i=1}^q D_{ii}^{-\frac{\alpha_i ( \mathcal{D})}2} 
	\end{align}
	\noindent
	for every $(D,L) \in \Theta_{ \mathcal{D}}$. 
	
	 The class of densities $\pi_{U,   \alpha( \mathcal D)}^{\Theta_{ \mathcal{D}}}$ form a 
	conjugate family of priors for the Gaussian DAG model $ \mathcal{N}( \mathcal{D})$. In particular, as indicated in \citep{BLMR:2016}, we have the following result that gives the MAP (maximum a posteriori) estimate for our Cholesky parameter.
	\begin{proposition} \label{prop1}
	If the prior on $(D,L) \in \Theta_{ \mathcal{D}}$ is $\pi_{U,   
			\alpha( \mathcal D)}^{\Theta_{ \mathcal{D}}}$ and $  {Y}_1, \ldots,   {Y}_n$ are 
		independent, identically distributed $N_p(  {0},(L^T)^{-1}DL^{-1})$ random vectors, 
		then the posterior distribution of $(D,L)$ is $\pi_{\tilde{U},\tilde{   \alpha}( \mathcal 
			D)}^{\Theta_{ \mathcal{D}}}$ with posterior mode $(\hat{D}, \hat{L})$ satisfying,
		\begin{equation} \label{posterior_mode}
		\hat{D}_{ii} = \frac{\tilde{U}_{i|pa_i( \mathcal D)}}{\alpha_i+n}, \quad \hat{L}_{ \mathcal D.i}^> = -[\tilde{U}_ \mathcal D^{>i}]^{-1}\tilde{U}_{ \mathcal D.i}^>,
		\end{equation}
		 where $S = \frac1n\sum_{i=1}^n  {Y}_i  {Y}_i^T$ 
		denotes the sample covariance matrix, $\tilde{U} = U + nS$, $\tilde{U}_{i|pa_i( \mathcal D)} = \tilde{U}_{ii} - (\tilde{U}_{ \mathcal D \cdot i}^>)^T 
		(\tilde{U}_ \mathcal D^{>i})^{-1} \tilde{U}_{ \mathcal D \cdot i}^>$, and $\tilde{   \alpha}
		( \mathcal D) = (n+\alpha_1( \mathcal D), \ldots, n+\alpha_p( \mathcal D))$. 
	\end{proposition}
\noindent
	Therefore, for any given DAG, Proposition \ref{prop1} yields the posterior mode for the Cholesky parameter and hence, the precision matrix $\Omega$. In particular, if the DAG imposed on the DAG-Wishart distribution is the true DAG, Section \ref{sec:posterior} provides the results for the posterior convergence rate.
		
	\section{Bayesian approach for precision matrix estimation} \label{sec:model}
	Let $\bm Y = \left({\bm Y}_1, {\bm Y}_2, \cdots, {\bm Y}_n \right)^T \in \mathbb{R}^{n\times p}$ be the observed data. The closed form for the posterior mode specified in (\ref{posterior_mode}) is convenient. However, it requires the underlying DAG and accordingly, the parent ordering to be given, which is rather problematic in real applications, especially the case when the ordering of variables and the conditional independence between variables are both unknown. We therefore propose a Bayesian approach to combine the graph selection procedures in \citep{CKG:2017} and permutations techniques in \citep{Kang:2017} to address this uncertainty of ordering issue and gain the flexibility to ensemble the multiple estimates for the Cholesky parameters for each ordering respectively.
	
	First we define a mapping function 
	\begin{equation}
	\sigma: \{1,2,\ldots, p\} \rightarrow \{1,2,\ldots, p\},
	\end{equation}
 such that $(\sigma(1), \sigma(2), \ldots, \sigma(p))$ represents a new permutation of $(1,2,\ldots, p)$. Define the corresponding permutation matrix $P_{\sigma}$ as follows. For each column in $P_{\sigma}$, all the entries in the $j$th column are set at 0 except the entry at the $\sigma(j)$th row equals to 1. Denote the new data matrix after permutation as$$\bm Y_\sigma = \bm Y P_\sigma,$$ and let $$S_\sigma = \frac 1 n \bm Y_\sigma^T \bm Y_\sigma$$ denote the new sample covariance matrix.

	In order to obtain the DAG-constraint posterior mode for the precision matrix with respect to this new data matrix $\bm Y_{\sigma}$, we must first acquire an appropriate estimate for the underlying DAG. The idea of hard thresholding the Cholesky factor of the sample covariance matrix to estimate the DAG structure has been implemented in \citep*{Bickel:Levina:2008, bickel:2008:thres, CKG:2017}. In particular, \citet*{CKG:2017} illustrate the graph selection consistency under the DAG-Wishart priors though the following procedure along with the justification of adopting this method.
	\begin{algorithm}[Estimate of the underlying DAG] \label{DAG_estimate}
		\item[Step 1:] Generate graphs by thresholding the modified Cholesky factor of $(S_\sigma + 0.1 I)^{-1}$ to get a sequence of $3000$ additional graphs, and search around 
		all the above graphs using Shotgun Stochastic Search to generate even more candidate graphs.
		\item[Step 2:] Random partition the original data set $\bm Y_\sigma$ of $n$ observations into $10$ equal sized subsets. Each time a single subset is excluded, and the remaining $9$ subsets are used as our new sample. Repeat Step 1 for each new sample.
		\item[Step 3:] Compute the log posterior probabilities for all cadidate graphs and select the graph $\hat{ \mathcal D}$ with the highest probability.
	\end{algorithm}
\noindent
\begin{remark}
	The numbers in Algorithm \ref{DAG_estimate} such as the length of sequence and the number of subsets can be modified depending on computational resources. Note that in Steps 2 and 3, due to the conjugacy of the DAG-Wishart distribution, the (marginal) posterior DAG probabilities can be computed as
	\begin{align} \label{m3}
	\pi({\mathcal{D}}| {Y}) \propto\frac{\gamma_{\mathcal{D}}(U+nS_\sigma,n+ {\alpha}(\mathcal D))}{\gamma_{\mathcal{D}}(U, {\alpha}(\mathcal D))}.
	\end{align}
\end{remark}
Now that we have the estimated DAG $\hat{ \mathcal D}$ for the new data matrix $\bm Y_\sigma$, the natural question arises as how to regenerate the modified Choleksy parameter and obtain the appropriate estimate for the true precision matrix. We therefore propose to use the MAP (maximum a posteriori) estimate given in Proposition \ref{prop1} with the following algorithm under permutation $\sigma$.
\begin{algorithm}[MAP estimate of $\Omega$] \label{MAP_estimate}
	\item[Step 1:] For $i = 1, 2, \ldots, p$, compute $d_{i} = \frac{\tilde{U}_{i|pa_i(\hat { \mathcal D})}}{\alpha_i+n}, l_i = -[\tilde{U}_{\hat { \mathcal D}}^{>i}]^{-1}\tilde{U}_{\hat { \mathcal D}.i}^>$, where $\tilde{U} = nS_\sigma + U$. Set $d_i= 0$ and $l_i = 0$, whenever $pa_i(\hat { \mathcal D}) = 0$.
	\item[Step 2:] Reconstruct Cholesky parameter $(\hat{D}_\sigma,\hat{L}_\sigma) \in \Theta_{\hat { \mathcal{D}}}$ satisfying $\left(\hat{L}_\sigma\right)_{\hat{ \mathcal D}.i}^> = l_i$ and $\hat{D}_\sigma = \mbox{diag}\left\{d_1,d_2,\ldots, d_p\right\}$.
	\item[Step 3:] Set $\hat \Omega_\sigma = \hat L_\sigma\hat D_\sigma^{-1} \hat L_\sigma^T$.
\end{algorithm}
\noindent
By transforming $\hat D_\sigma$, $\hat L_\sigma$ and $\hat \Omega_\sigma$ to the original order, we can thereby estimate $D, L$ and $\Omega$ with 
\begin{align} \label{perm_L_D}
\hat L = P_\sigma \hat L_\sigma P_\sigma^T, \quad \hat D = P_\sigma \hat D_\sigma P_\sigma^T
\end{align}
and 
\begin{align} \label{perm_Omega}
\hat \Omega = \hat L \hat D^{-1} \hat L^T = P_\sigma \hat L_\sigma P_\sigma^T(P_\sigma \hat D_\sigma P_\sigma^T)^{-1}P_\sigma \hat L_\sigma^T P_\sigma^T = P_\sigma \hat \Omega_\sigma P_\sigma^T.
\end{align}

Now suppose we generate $K$ permutations denoted as $\sigma_1, \sigma_2, \ldots, \sigma_K$. For each permutation $\sigma_k$ $(1 \le k  \le K)$, Algorithm \ref{MAP_estimate}, (\ref{perm_L_D}) and (\ref{perm_Omega}) yield the corresponding estimates $\hat{L}_k$, $\hat{D}_k$ and $\hat{\Omega}_k$. Naturally, there are two ways for integrating these estimates. The first approach is to average $\Omega_1, \Omega_2, \ldots, \Omega_K$ and let $\bar \Omega = \frac 1 K \sum_{k=1}^{K} \hat \Omega_k$ be our final estimate for the true precision matrix. However, this average estimate performs poorly, as the estimation error of each $\hat \Omega_k$ is already aggregated by the error in both $\hat{L}_k$ and $\hat D_k$. 
We therefore propose to use the average estimates of both $\hat L_k$'s and $\hat D_k$'s to construct the ensemble estimate for $\Omega$. In particular, denote 
\begin{equation} \label{bar_L_D}
\bar{L} = \frac 1 K \sum_{k=1}^{K} \hat{L}_k, \quad \bar{D} = \frac 1 K \sum_{k=1}^{K} \hat{D}_k,
\end{equation}
and estimate $\Omega$ with 
\begin{align} \label{final_Omega}
\check{\Omega} = \bar L \bar D^{-1} \bar L^T.
\end{align}
\noindent
In Section \ref{sec:simulation}, we will see this estimate can have the advantage of averaging the variability and better recover the individual entry values of the true precision matrix. 

However, $\check \Omega$ does not necessarily carry out the true sparse pattern encoded in possibly very parse true inverse covariance matrix. Hence, we utilize the following hard thresholding procedure to encourage the sparsity in $\check \Omega$. For any given thresholding value $\tau$, construct the corresponding sparse matrix $\bar L_\tau$ based on $\bar L$ in (\ref{bar_L_D}) as follows. For $1 \le i,j \le p$, 
\[(\bar{L}_\tau)_{ij} = 
\begin{cases}
\bar L_{ij}, & |\bar L_{ij}| > \tau \\
0, & |\bar L_{ij}| \le \tau.
\end{cases} 
\]
Estimate $\Omega$ with $\check{\Omega}_\tau = \bar L_\tau \bar D^{-1} \bar L_\tau^T$. If we vary the thresholding value $\tau$ on a grid, the best thresholding value $\tau_b$ is selected, which minimizes the ``BIC"-like measure defined as \begin{equation} \label{BIC}
BIC(\tau) = ntr(S\check{\Omega}_\tau) - n\log|\check{\Omega}_\tau| + \log n * E,
\end{equation}
where $E$ denotes the total numbers of non-zero entries in $\bar{L}_\tau$ and $\check{\Omega}_\tau = \bar L_\tau \bar D^{-1} \bar L_\tau^T.$ See examples in \citep{CKG:2017, KORR:2017, Shojaie:Michailidis:2010, Kang:2017} for justifications of the ``BIC-like criterion. Note that the estimate $\check \Omega_{\tau_b}$ possesses a much more sparse structure compared to $\check \Omega$ in (\ref{final_Omega}) via the thresholding step, and could possibly better reveal the true sparsity pattern in the underlying precision matrix. Now we are ready to present the following algorithm for our proposed Bayesian approach for estimating precision matrix.
 \begin{algorithm} \label{final_algorithm}
 	\item Generate $K$ permutations $\sigma_1, \sigma_2, \ldots, \sigma_K$ and obtain the corresponding data matrix and sample covariance matrix $\bm Y_{\sigma_k}$, $S_{\sigma_k}$. For $1 \le k \le K$, do step 1-3.
 	\item[Step 1:] Implement Algorithm \ref{DAG_estimate} and obtain the estimated DAG $ \mathcal D_{\sigma_k}.$
 	\item[Step 2:] Given $ \mathcal D_{\sigma_k}$, obtain $\hat{L}_{\sigma_k}$, $\hat{D}_{\sigma_k}$, $\hat{\Omega}_{\sigma_k}$ via Algorithm \ref{MAP_estimate}.
 	\item[Step 3:] Set $\hat L_k = P_{\sigma_k} \hat L_{\sigma_k} P_{\sigma_k}^T, \quad \hat D_k = P_{\sigma_k}  \hat D_{\sigma_k}  P_{\sigma_k} ^T$.
 	\item[Step 4:] Obtain the average estimates $\bar{L} = \frac 1 K \sum_{k=1}^{K} \hat{L}_k, \bar{D} = \frac 1 K \sum_{k=1}^{K} \hat{D}_k,$ and $\check{\Omega} = \bar L \bar D^{-1} \bar L^T.$
 	\item[Step 5:] Vary the thresholding values on a grid and select $\bar L_{\tau_b}$ according to the measure in (\ref{BIC}).
 	\item[Step 6:] Set $\check{\Omega}_{\tau_b} = \bar L_{\tau_b} \bar D^{-1} \bar L_{\tau_b}^T$ as the estimate for $\Omega$.
 \end{algorithm}
\begin{remark}
	We would like to point out that the hard thresholding procedure in Step 3 and 4 is not necessarily required, if one is merely interested in the estimation rather than the sparsity recovery of the precision matrix. As we will see in Section \ref{sec:simulation}, no significant difference is observed in performance between the estimate $\check{\Omega}$ obtained in Step 4 and the final estimate $\check{\Omega}_{\tau_b}$ under certain settings. Hence, $\check{\Omega}$ can also serve as an alternate estimate in practice.
\end{remark}
\section{Simulation studies} \label{sec:simulation}
In this section, we illustrate the potential advantage of utilizing the proposed Bayesian approach through simulation studies. We fix our number of observations $n = 100$. We then consider different combinations of $(n,p)$ with $p$ ranging from smaller than $30$ to $200$, such that $p$ varies from smaller than $n$ to larger than $n$. Next, for each fixed $p$, a $p \times p$ positive definite matrix $\Omega_0$ is constructed. In particular, we consider the following five cases of $\Omega_0$, which are also considered in \citep*{Kang:2017}.
\begin{description}
	\item [$Case\mbox{ } 1$:] $\Omega_0$ has a banded structure such that the main diagonal entries equal to 1, while the first sub-diagonal entries equal to 0.5 and second sub-diagonals 0.3.
	\item [$Case \mbox{ }2$:] $\Sigma_0 = \Omega_0^{-1}$ is autoregressive correlated such that $(\Sigma_0)_{ij} = 0.5^{|i-j|},$ for all $1 \le i \le j \le p$.
	\item [$Case \mbox{ }3$:] $\Omega_0$ has $3\%$ sparsity. For each fixed $p$, we start from a $p \times p$ identity matrix. Then, we randomly choose $3\%$ of the lower triangular entries and set the values to be randomly drawn from $\mbox{Unif}(0,1)$. We refer to this matrix as our true Cholesky factor $L_0$ and set $\Omega_{0} = L_0L_0^T$.
	\item [$Case \mbox{ }4$:] $\Omega_0$ has a $10\times 10$ compound structure matrix on the left top with diagonal elements 1 and others 0.5. Other entries of $\Omega_{0}$ are set to be zero, except for all the diagonals taken to be 1.
	\item [$Case \mbox{ }5$:] $\Omega_0$ is generated by randomly permuting rows and corresponding columns of the precision matrix in Case 4.
\end{description}
Next, for every case of $\Omega_0$, we generate 
$n$ i.i.d. observations from the $N(0_p, \Omega_0)$ distribution. We then estimate the true precision matrix $\Omega_0$ using both Bayesian and frequentist procedures outlined below. 
\begin{description}
	\item[$DAGW.BIC$:] Set the hyperparameters for our DAG-Wishart distribution in (\ref{a3}) as $U = I_p$, $\alpha_i(\mathcal D) = \nu_i(\mathcal D) + 10$ for $i = 1,2,\ldots,p$, and the total number of permutations $K=100$. Obtain our estimate $\check{\Omega}_{\tau_b}$ via Algorithm \ref{final_algorithm}.
	\item[$DAGW$:] Under the same hyperparameter setting, obtain the estimate $\check{\Omega}$ via Algorithm \ref{final_algorithm}, but without the step of hard thresholding, i.e., ${\tau_b} = 0$.
	\item[$MLE$:] For each permutation, replace MAP in \ref{MAP_estimate} with the $\mathcal D_{\sigma_k}$-constrained maximum likelihood estimate. Then implement step 2--4 in Algorithm \ref{final_algorithm} to obtain the ensemble MLE estimate. 
	\item [$BAYES$:] Under the same hyperparameter setting, implement Algorithm \ref{DAG_estimate} and \ref{MAP_estimate} with the original data without permutations.
	\item[$MCD.BIC$:] Implement the improved MCD approach based on penalized likelihood approaches in~\citep{Kang:2017} with BIC-based tuning. 
	\item [$PC-DAG$:] Implement the PC-algorithm based approach for inverse covariance estimation introduced in \citep*{PC:2009} encoded in R package ``pcalg" with suggested tuning parameters.
\end{description}
We then compare the estimation performance between these six methods under the following five different losses. Stein's loss is a commonly used loss function given by $$L_1(\hat{\Omega}, \Omega_0) = tr(\hat{\Omega}\Omega_0^{-1}) - \log(det(\Omega_0\Omega_0^{-1})) - p,$$ where $\hat{\Omega}$ represents the estimator of the true precision matrix $\Omega_0$. The modified mean absolute error loss and mean squared error loss, restricted to the functionally independent elements of the true precision matrix are defined as,
$$L_2(\hat{\Omega}, \Omega_0) = \sum_{i = 1}^q\sum_{j \in pa_i( \mathcal D_0)}\lvert(\Omega_0)_{ji} - \hat \Omega_{ji}\rvert, \quad L_3(\hat{\Omega}, \Omega_0) = \sum_{i = 1}^q\sum_{j \in pa_i( \mathcal D_0)}\left((\Omega_0)_{ji} - \hat \Omega_{ji}\right)^2,$$
where $ \mathcal D_0$ represents the true underlying DAG implying the sparsity pattern in $\Omega_0$.
We also adopt the following two measures of accuracy: the general mean absolute error and mean squared error given by
$$L_4(\hat{\Omega}, \Omega_0) = \sum_{i = 1}^q\sum_{j = 1}^q\lvert(\Omega_0)_{ji} - \hat \Omega_{ji}\rvert, \quad L_5(\hat{\Omega}, \Omega_0) = \sum_{i = 1}^q\sum_{j = 1}^q\left((\Omega_0)_{ji} - \hat \Omega_{ji}\right)^2. $$        
The summary of our results are presented in Table \ref{table:MA} to Table \ref{table:diag_perm} corresponding to five different structures of the true precision matrix respectively. The five losses for six methods averaged over 20 repetitions and their corresponding standard errors (in parenthesis) for different approaches are shown in the tables. For each loss, the lowest averages among all the methods are highlighted. We can see from the results, under different scenarios of the true precision matrix, our proposed method DAGW.BIC and DAGW outperform other frequentist methods and Bayesian methods under almost all the five measures. In particular, when the dimension $p$ increases, our proposed permutation method with DAG-Wishart prior can sustain the higher dimension and achieve much better and more stable estimation results. 

\begin{figure}[htbp]
		\begin{subfigure}[t]{.24\textwidth}
			\centering
			\includegraphics[width=.8\linewidth]{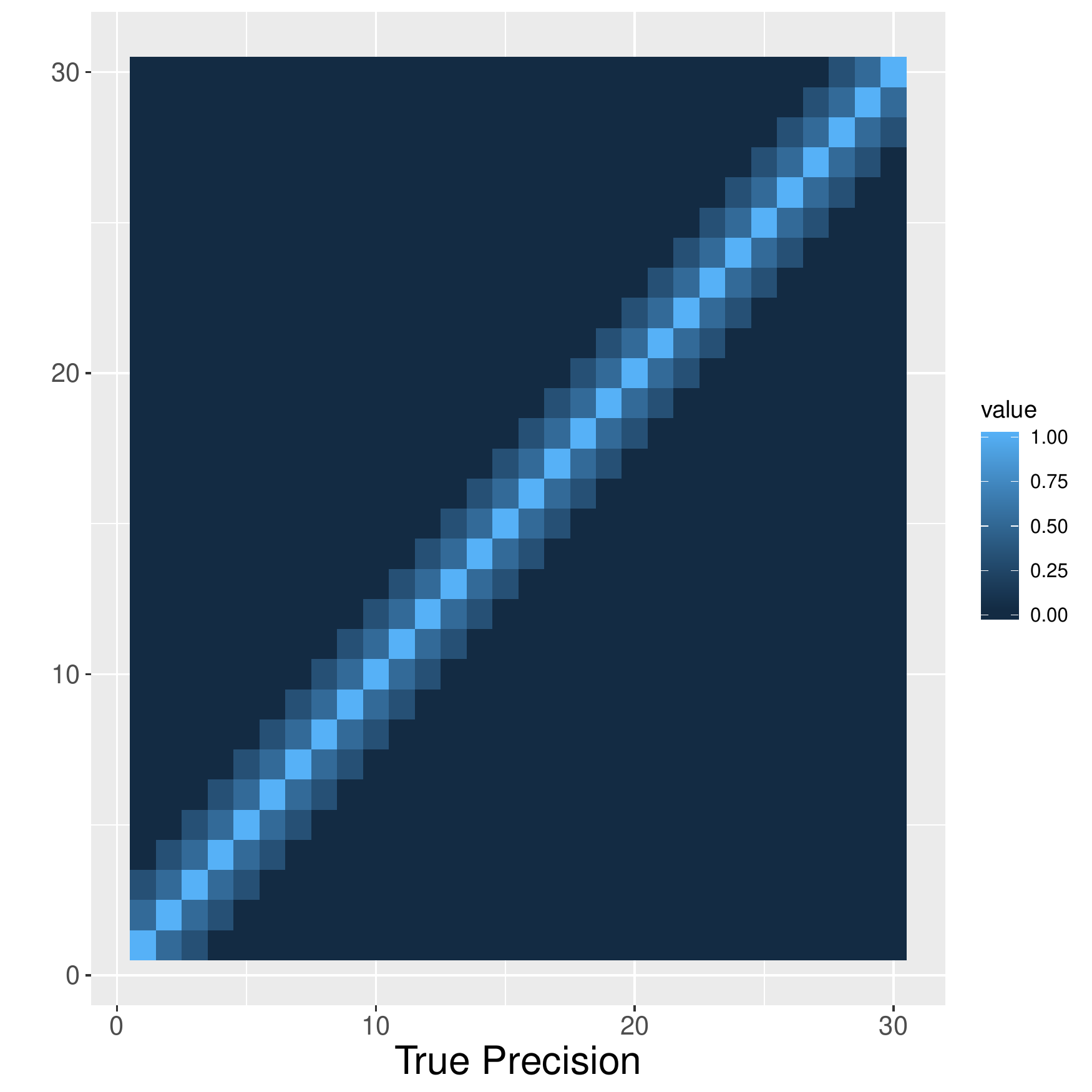}
			\caption{True precision}
		\end{subfigure}%
		\begin{subfigure}[t]{.24\textwidth}
			\centering
			\includegraphics[width=.8\linewidth]{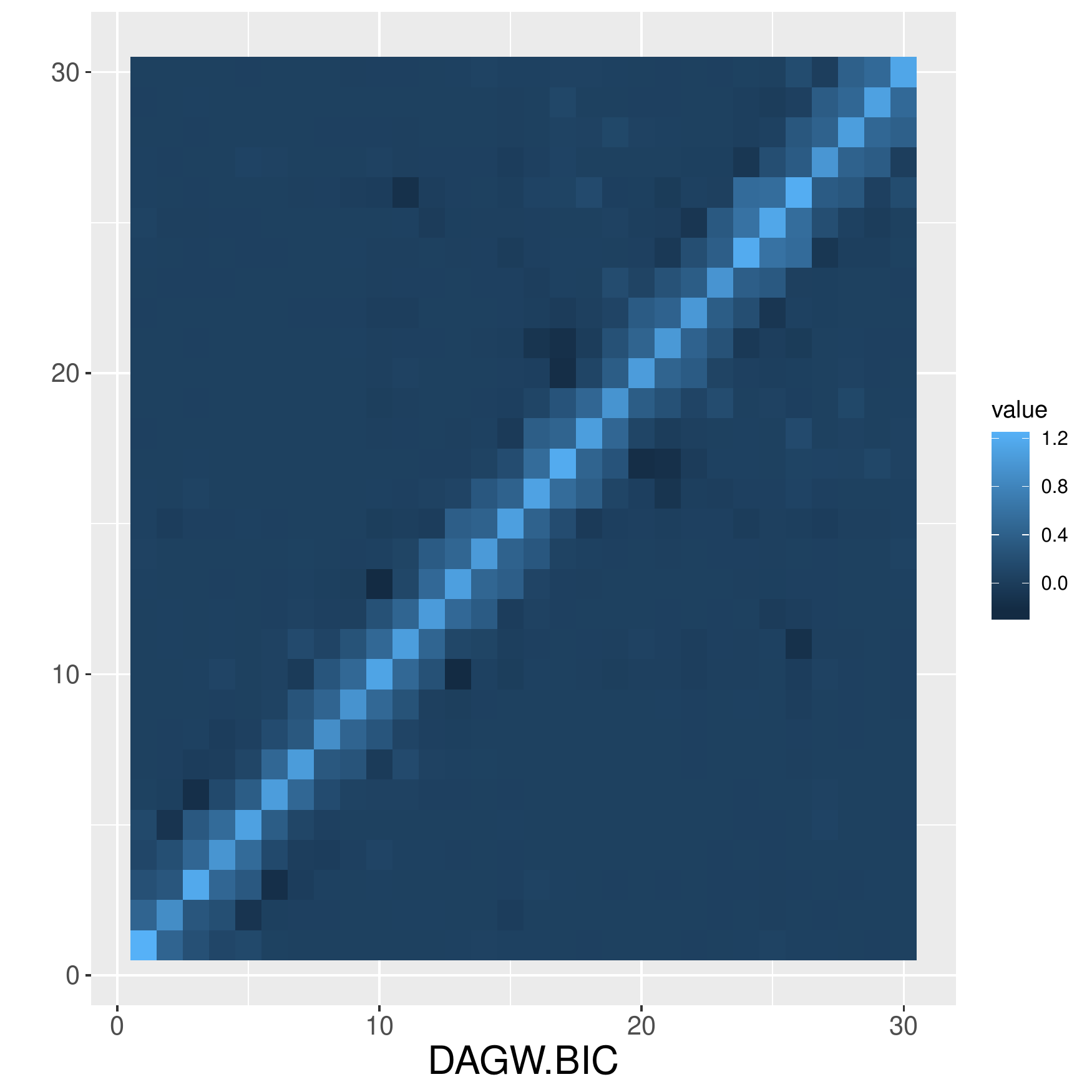}
			\caption{DAGW.BIC}
		\end{subfigure}
		\begin{subfigure}[t]{.24\textwidth}
			\centering
			\includegraphics[width=.8\linewidth]{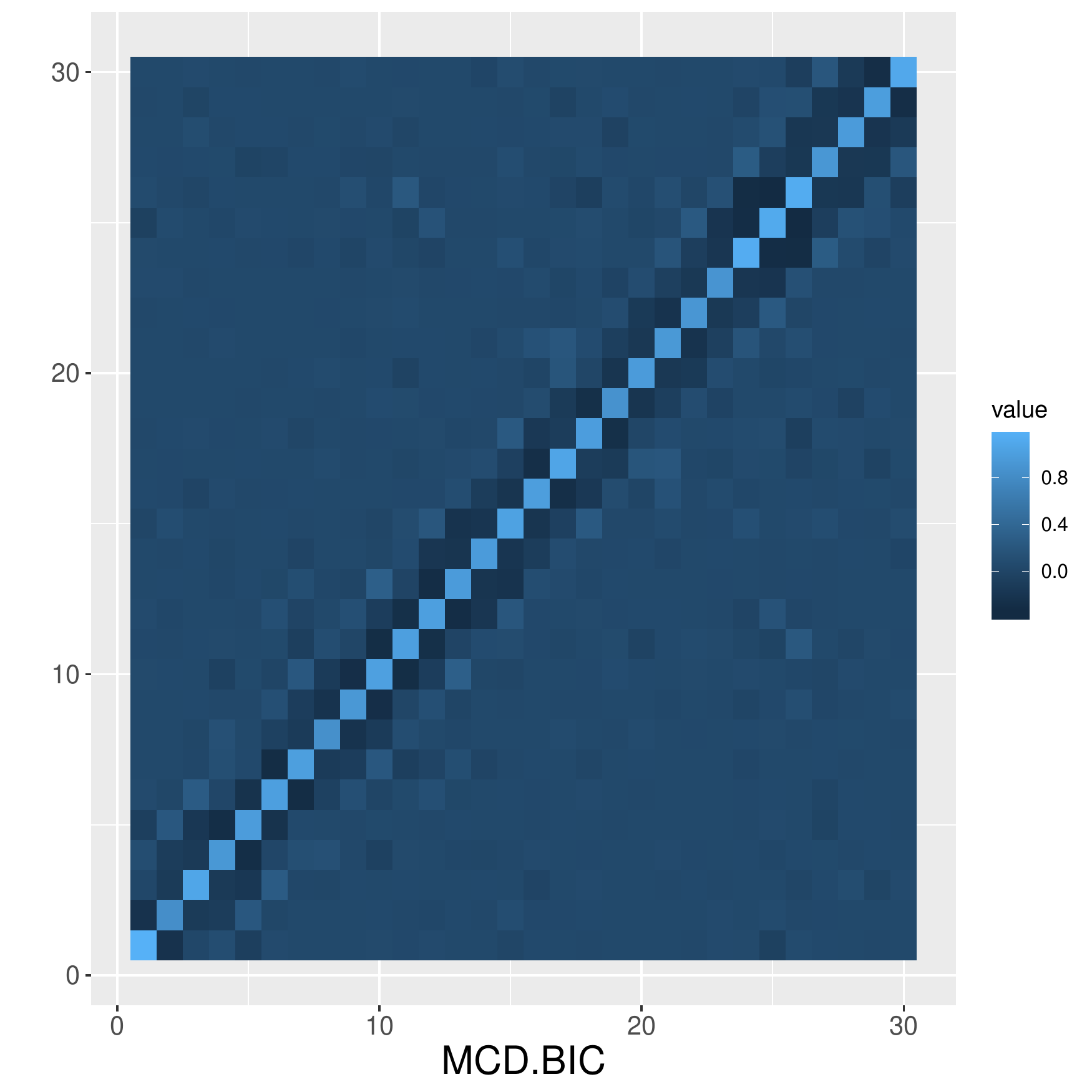}
			\caption{MCD.BIC}
		\end{subfigure}
		\begin{subfigure}[t]{.24\textwidth}
			\centering
			\includegraphics[width=.8\linewidth]{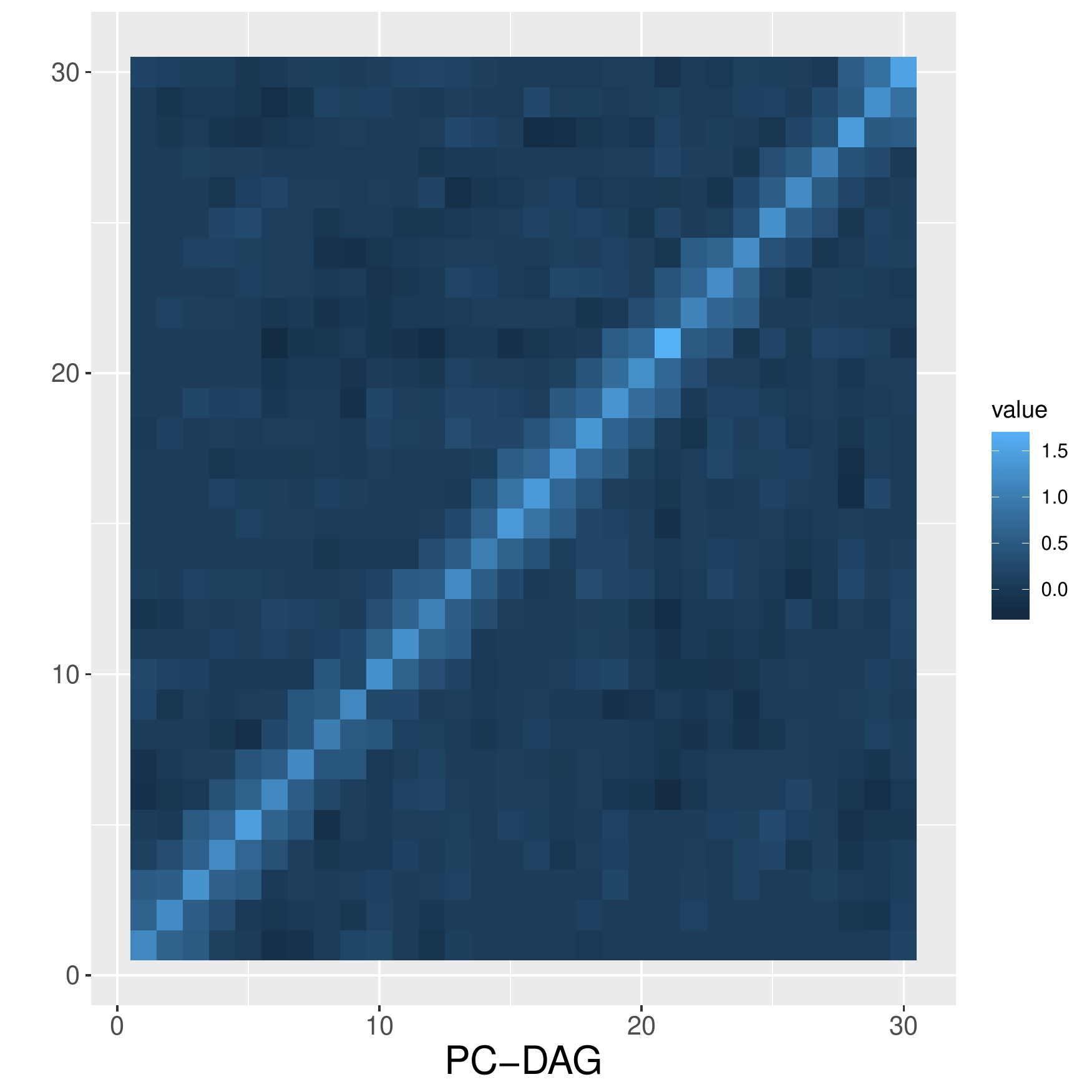}
			\caption{PC-DAG}
		\end{subfigure}
		\caption{Heatmap comparison of estimates when $p = 30$}
		\label{figure:p30}
	\end{figure}

\begin{figure}[htbp]
		\begin{subfigure}[t]{.24\textwidth}
			\centering
			\includegraphics[width=.8\linewidth]{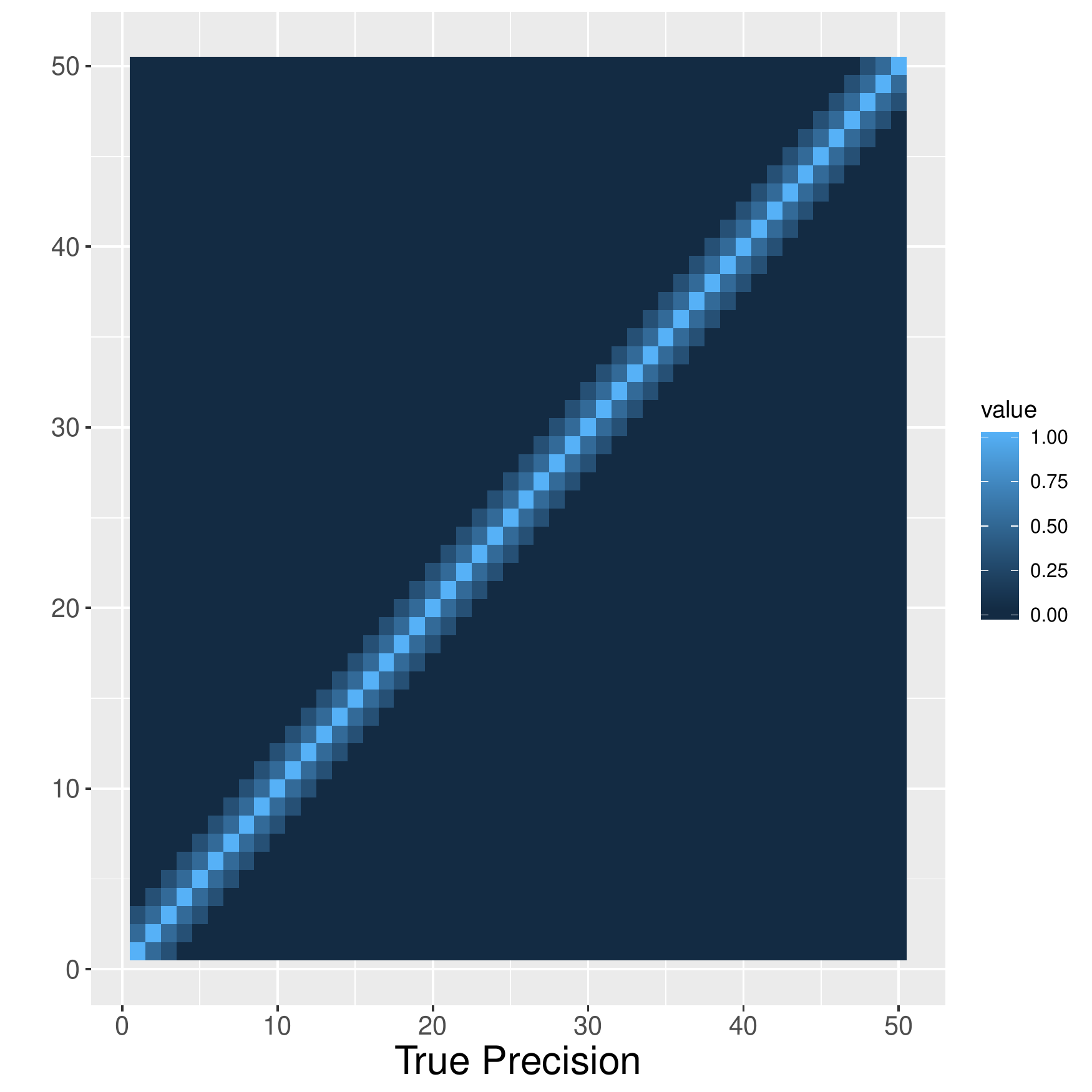}
			\caption{True precision}
		\end{subfigure}%
		\begin{subfigure}[t]{.24\textwidth}
			\centering
			\includegraphics[width=.8\linewidth]{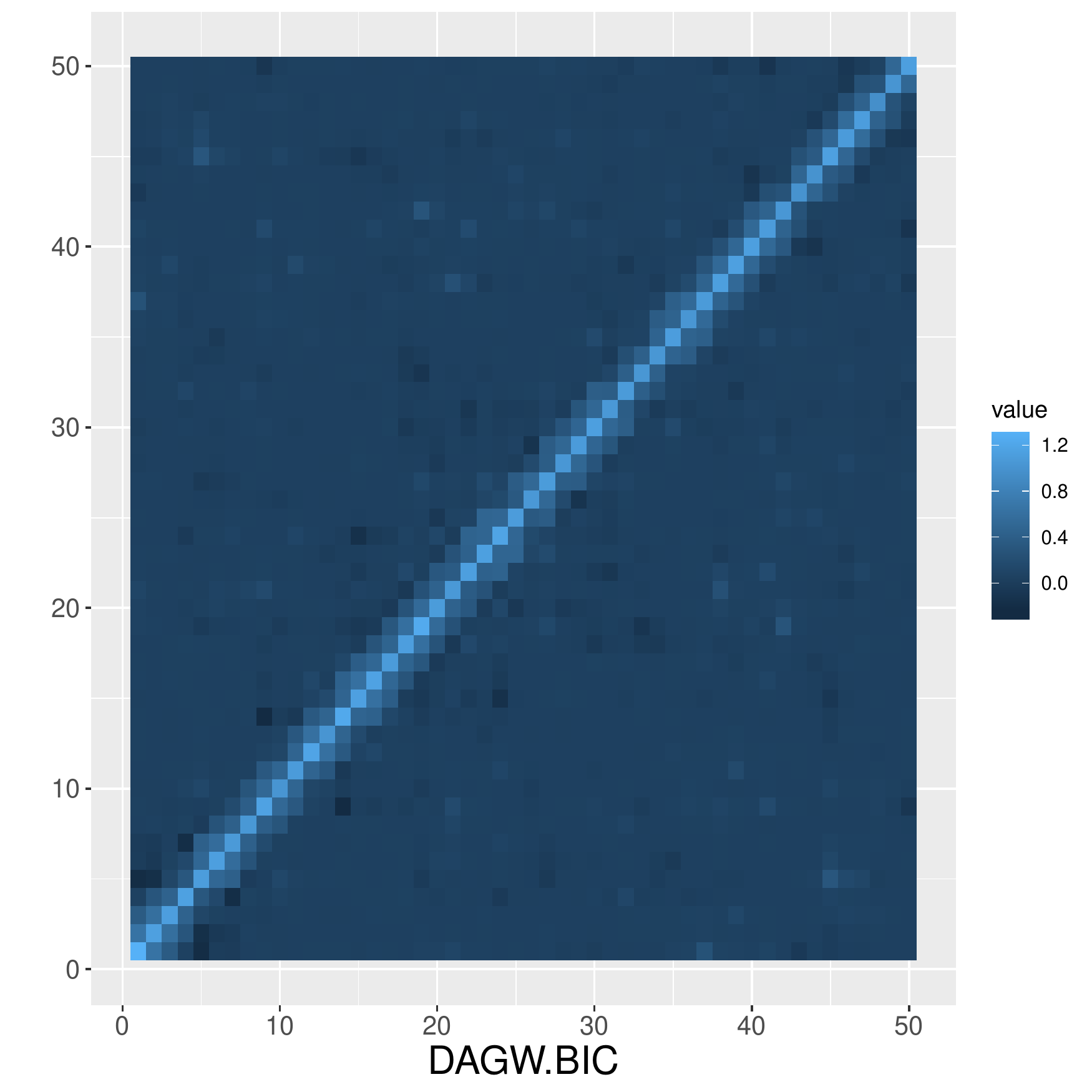}
			\caption{DAGW.BIC}
		\end{subfigure}
		\begin{subfigure}[t]{.24\textwidth}
			\centering
			\includegraphics[width=.8\linewidth]{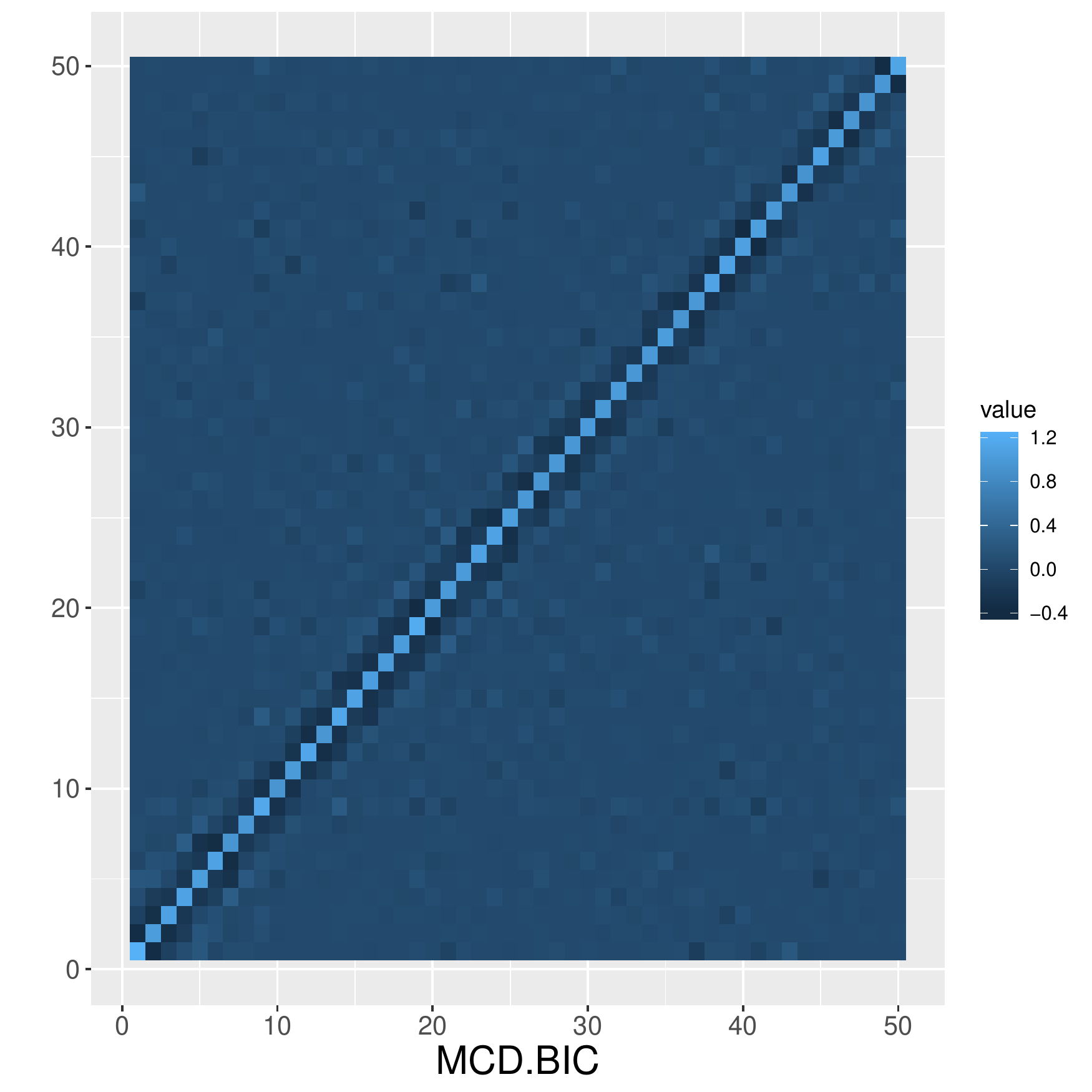}
			\caption{MCD.BIC}
		\end{subfigure}
		\begin{subfigure}[t]{.24\textwidth}
			\centering
			\includegraphics[width=.8\linewidth]{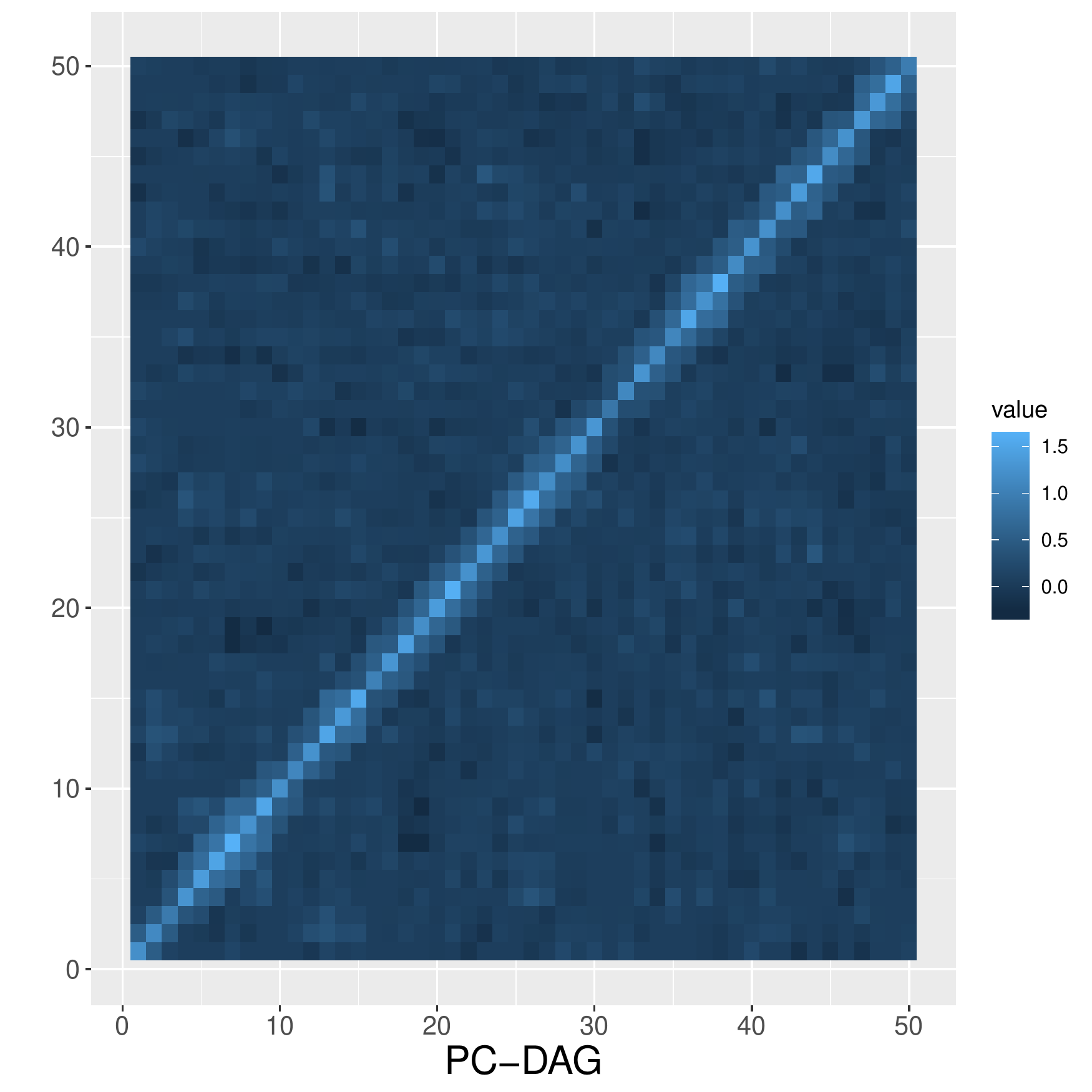}
			\caption{PC-DAG}
		\end{subfigure}
		\caption{Heatmap comparison of estimates when $p = 50$}
		\label{figure:p50}
	\end{figure}

\begin{figure}[htbp]
		\begin{subfigure}[t]{.24\textwidth}
			\centering
			\includegraphics[width=.8\linewidth]{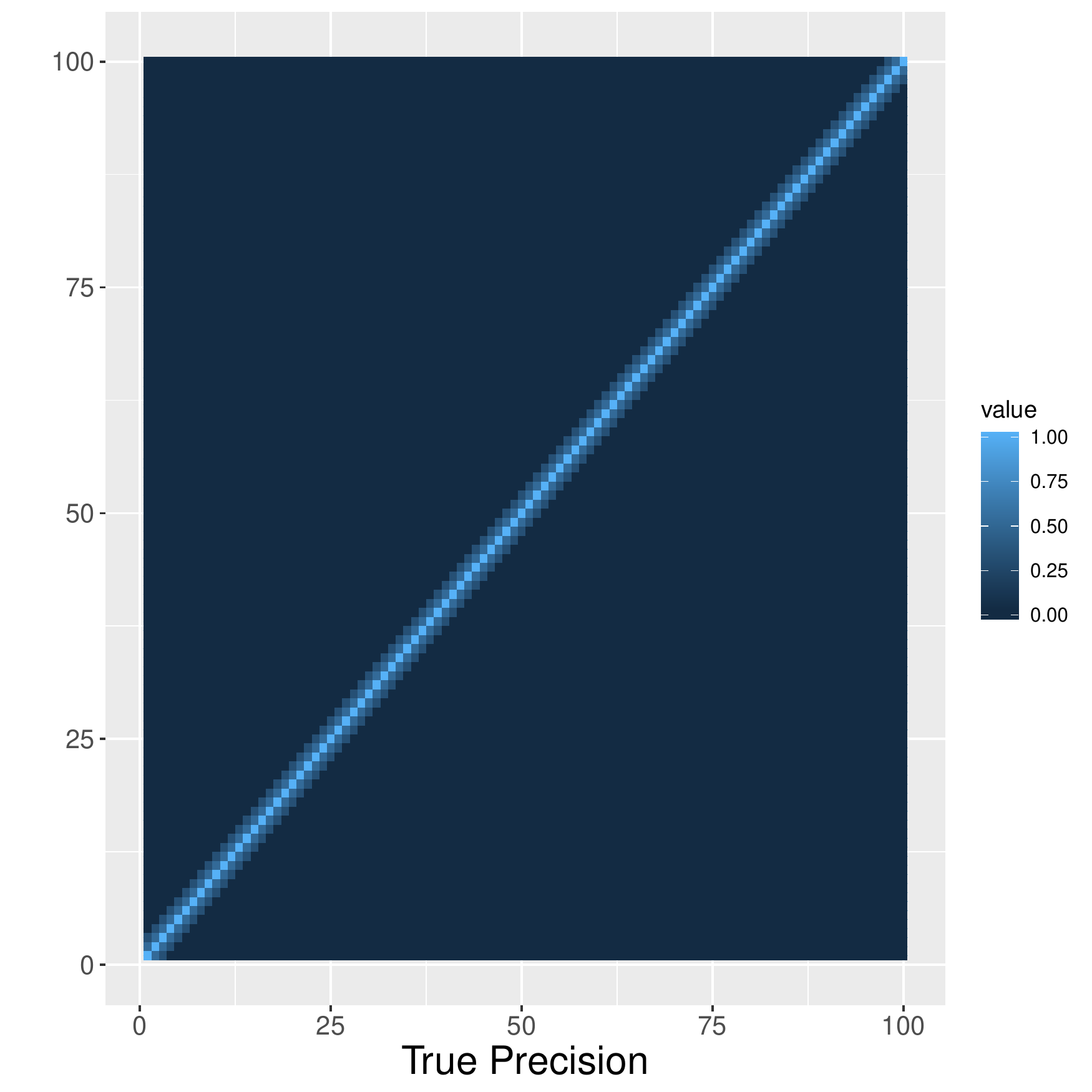}
			\caption{True precision}
		\end{subfigure}%
		\begin{subfigure}[t]{.24\textwidth}
			\centering
			\includegraphics[width=.8\linewidth]{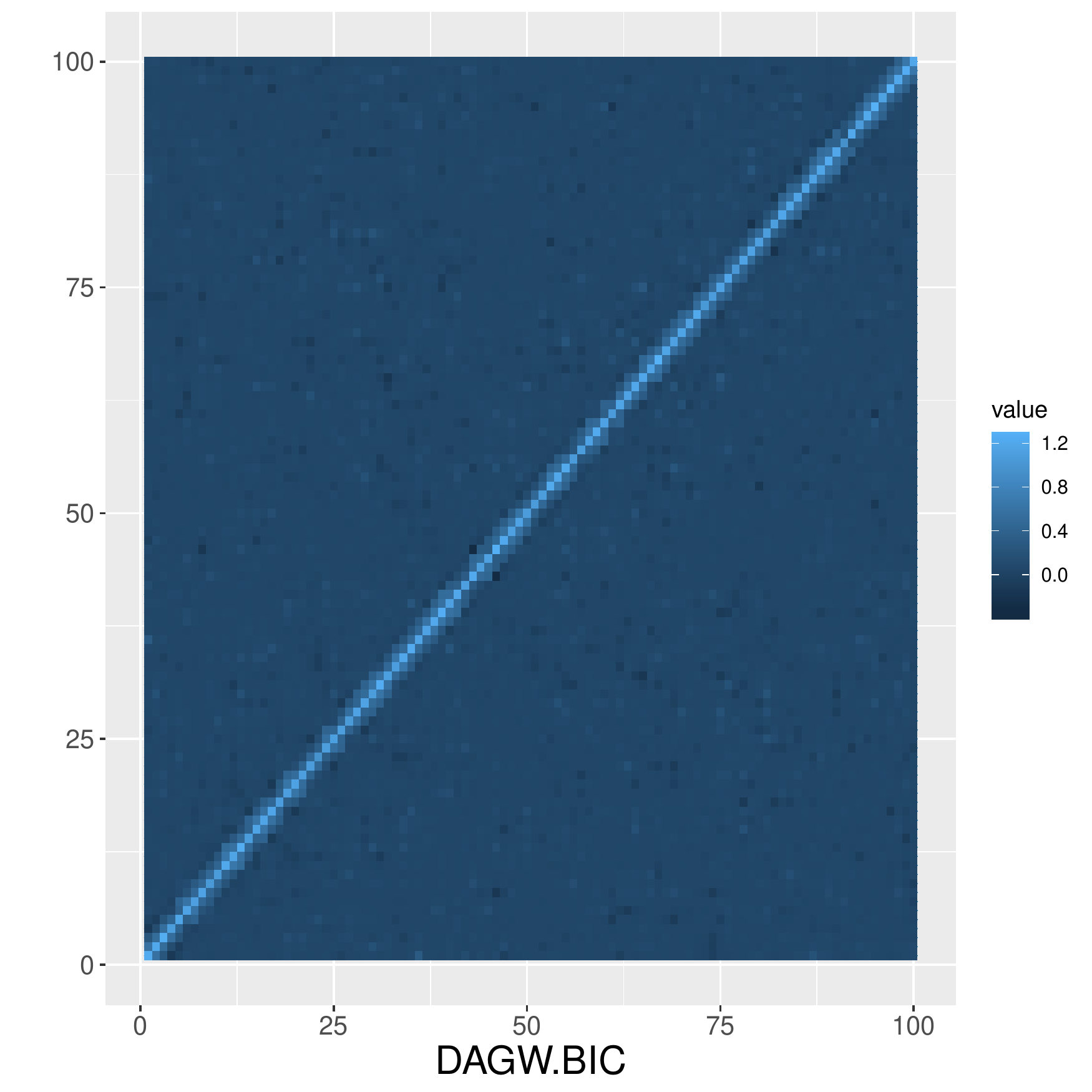}
			\caption{DAGW.BIC}
		\end{subfigure}
		\begin{subfigure}[t]{.24\textwidth}
			\centering
			\includegraphics[width=.8\linewidth]{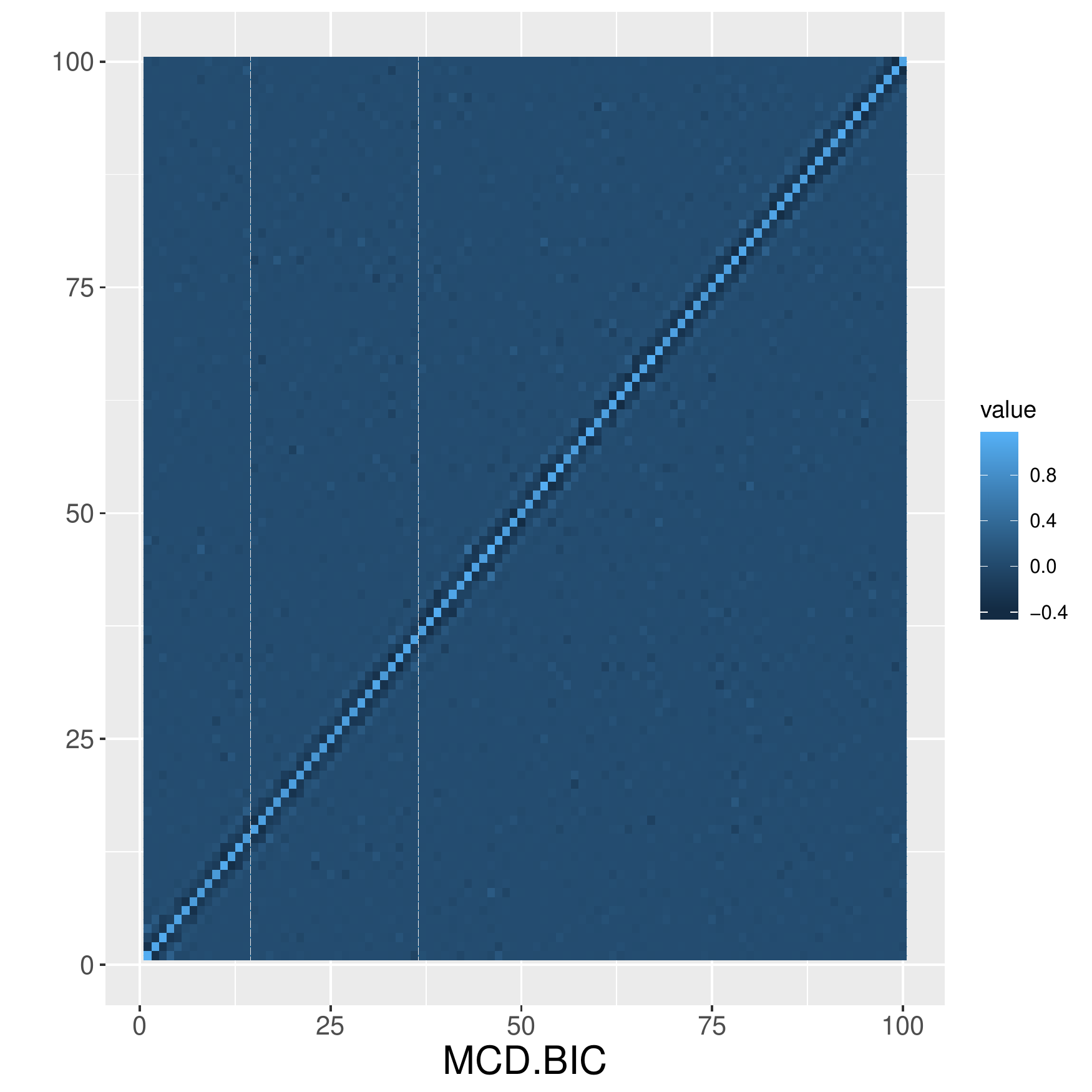}
			\caption{MCD.BIC}
		\end{subfigure}
		\begin{subfigure}[t]{.24\textwidth}
			\centering
			\includegraphics[width=.8\linewidth]{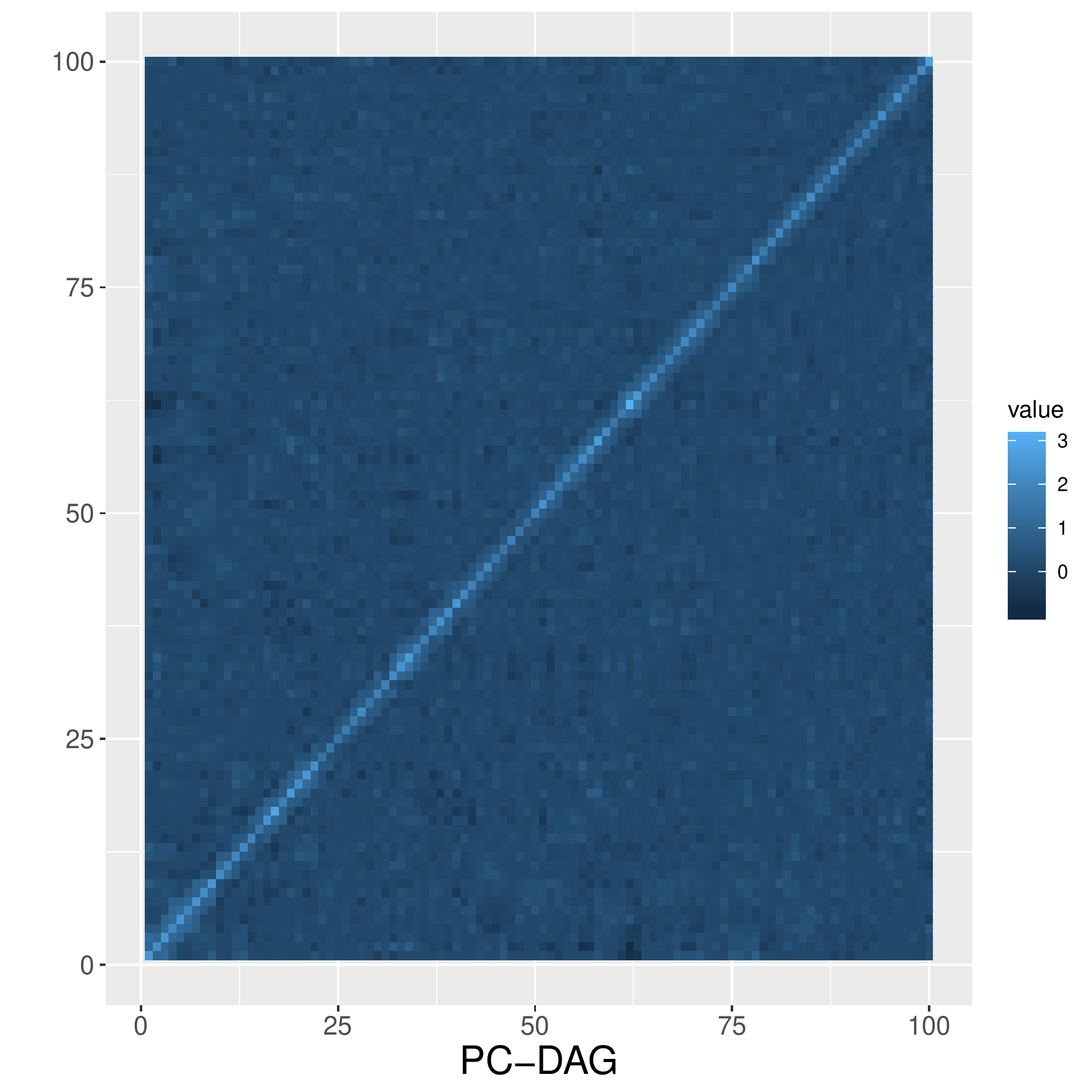}
			\caption{PC-DAG}
		\end{subfigure}
		\caption{Heatmap comparison of estimates when $p = 100$}
		\label{figure:p100}
	\end{figure}

To better visualize the simulation results, we plot the heatmap comparison between the true precision matrix and different estimators under different values of $p$ when the true precision matrix has a banded structure. In Figure \ref{figure:p30} to Figure \ref{figure:p100}, we can see among all three estimators, our proposed method can best recover the sparse structure of the true precision matrix. MCD method in \citep*{Kang:2017} fails to capture the structure for the first and second sub-diagonal entries, while the PC-DAG generates more false positives and obscures the true clear sparsity pattern. 
\begin{table}
	\begin{tabular}{cccccccc} \hline
		&    & DAGW.BIC                          & DAGW                              & MLE                               & BAYES      & MCD.BIC    &PC-DAG     \\ \hline
		$p=30$  & $L_1$ & \textbf{0.07 (0.01)} & 0.15 (0.01)        & 0.16 (0.01)                        & 0.33 (0.00) & 1.41 (0.05) & 0.18 (0.02)  \\
		& $L_2$ & \textbf{0.03 (0.01)} & 0.03 (0.01)                        & 0.04 (0.01)                        & 0.32 (0.00) & 0.80 (0.02) & 0.10 (0.01)  \\
		& $L_3$ & \textbf{0.18 (0.02)} & 0.19 (0.02)                        & 0.22 (0.02)                        & 0.76 (0.00) & 1.21 (0.02) & 0.38 (0.03)  \\
		& $L_4$ & \textbf{0.08 (0.01)} & 0.19 (0.02)                        & 0.24 (0.02)                        & 0.88 (0.01) & 1.67 (0.04) & 0.29 (0.03)  \\
		& $L_5$ & \textbf{0.75 (0.06)} & 1.11 (0.07)                          & 1.23 (0.08)                        & 2.00 (0.01) & 3.07 (0.06) & 1.29 (0.11)  \\ \hline
		$p=50$  & $L_1$ & \textbf{0.09 (0.01)} & 0.16 (0.01)                        & 0.18 (0.01)         & 0.33 (0.00) & 1.40 (0.03) & 0.20 (0.02)  \\
		& $L_2$ & \textbf{0.03 (0.00)} & 0.03 (0.00)                        & 0.03 (0.00)                       & 0.33 (0.00) & 0.78 (0.01) & 0.11 (0.01)  \\
		& $L_3$ & \textbf{0.18 (0.01)} & 0.18 (0.01)                        & 0.21 (0.01)                        & 0.78 (0.00) & 1.18 (0.01) & 0.39 (0.03)  \\
		& $L_4$ & \textbf{0.10 (0.01)} & 0.20 (0.01)                        & 0.25 (0.01)                        & 0.90 (0.01) & 1.64 (0.03) & 0.31 (0.04)  \\
		& $L_5$ & \textbf{1.00 (0.06)} & 1.38 (0.05)                        & 1.54 (0.06)                        & 2.04 (0.01) & 3.23 (0.08) & 1.49 (0.11)  \\ \hline
		$p=100$ & $L_1$ & \textbf{0.15 (0.01)} & 0.20 (0.01)                        & 0.23 (0.01)                        & 0.34 (0.00) & 1.33 (0.02) & 0.27 (0.01)  \\
		& $L_2$ & \textbf{0.03 (0.00)} & 0.03 (0.00)                        & 0.04 (0.00)                        & 0.33 (0.00) & 0.75 (0.01) & 0.11 (0.01) \\
		& $L_3$ & 0.20 (0.01)                        & \textbf{0.19 (0.01)} & 0.21 (0.01)                        & 0.79 (0.00) & 1.16 (0.01) & 0.40 (0.02)  \\
		& $L_4$ & \textbf{0.16 (0.01)} & 0.22 (0.01)                        & 0.29 (0.01)                        & 0.91 (0.01) & 1.55 (0.02) & 0.36 (0.02)  \\
		& $L_5$ & \textbf{1.83 (0.08)} & 2.11 (0.07)                        & 2.40 (0.08)                        & 2.06 (0.01) & 3.40 (0.08) & 1.87 (0.07) \\ \hline
		$p=150$ & $L_1$ & \textbf{0.19 (0.01)} & 0.23 (0.01)                        & 0.27 (0.01)                        & 0.34 (0.02) & 1.28 (0.02) & 0.31 (0.01)  \\
		& $L_2$ & 0.05 (0.01)                        & \textbf{0.04 (0.01)} & 0.04 (0.00)                        & 0.34 (0.00) & 0.71 (0.01) & 0.12 (0.01)  \\
		& $L_3$ & 0.25 (0.02)                        & \textbf{0.23 (0.01)} & 0.30 (0.01)                        & 0.79 (0.00) & 1.12 (0.02) & 0.42 (0.02)  \\
		& $L_4$ & \textbf{0.19 (0.01)} & 0.23 (0.01)                        & 0.30 (0.01)                        & 0.92 (0.01) & 1.51 (0.01) & 0.40 (0.02) \\
		& $L_5$ & {2.19 (0.05)} & 2.43 (0.05)                        & 2.75 (0.05)                        & 2.07 (0.01) & 3.47 (0.03) & \textbf{2.12 (0.07)} \\ \hline
		$p=200$ & $L_1$ & \textbf{0.22 (0.01)} & 0.31 (0.01)                        & 0.36 (0.01)                        & 0.34 (0.00) & 1.28 (0.03) & 0.35 (0.01)  \\
		& $L_2$ & 0.08 (0.01)                        & 0.07 (0.01)                        & \textbf{0.06 (0.00)} & 0.34 (0.01) & 0.71 (0.01) & 0.12 (0.01)  \\
		& $L_3$ & 0.34 (0.01)                        & \textbf{0.30 (0.01)} & 0.30 (0.01)                        & 0.79 (0.02) & 1.16 (0.02) & 0.43 (0.02)  \\
		& $L_4$ & \textbf{0.22 (0.01)} & 0.26 (0.01)                        & 0.32 (0.01)                        & 0.92 (0.01) & 1.48 (0.03) & 0.44 (0.02) \\
		& $L_5$ & \textbf{2.24 (0.01)} & 2.60 (0.02)                        & 2.87 (0.03)                        & 2.28 (0.01) & 3.84 (0.11) & 2.39 (0.06) \\ \hline
	\end{tabular}
\caption{The loss averages and standard errors (in parenthesis) of estimates when $\Omega_0$ has a banded structure.} \label{table:MA}
\end{table}

\begin{table} 
	\begin{tabular}{cccccccc}\hline
		\multicolumn{1}{l}{} & \multicolumn{1}{l}{} & DAGW.BIC                          & DAGW                              & MLE        & BAYES      & MCD.BIC    &PC-DAG      \\ \hline
		p=30  & $L_1$ & \textbf{0.05 (0.01)} & 0.14 (0.01) & 0.14 (0.01) & 0.29 (0.00)    & 0.88 (0.03) & 0.06 (0.01)\\
	& $L_2$ & {0.05 (0.01)} & 0.11 (0.01) & 0.10 (0.01)  & 0.43 (0.00)    & 1.15 (0.05) & \textbf{0.04 (0.01)} \\
	& $L_3$ & \textbf{0.19 (0.02)} & 0.32 (0.01) & 0.31 (0.01) & 0.64 (0.00)    & 1.04 (0.02) & 0.13 (0.03) \\
	& $L_4$ & \textbf{0.26 (0.03)} & 0.76 (0.03) & 0.73 (0.03) & 1.28 (0.03) & 2.43 (0.08) & 0.26 (0.05) \\
	& $L_5$ & \textbf{1.04 (0.05)} & 1.58 (0.04) & 1.56 (0.04) & 1.92 (0.03) & 3.11 (0.10) & 1.04 (0.12)\\	\hline
	p=50  & $L_1$ &\textbf{0.05 (0.00)} & 0.15 (0.01) & 0.15 (0.01) & 0.29 (0.00)    & 0.91 (0.04) & 0.08 (0.01)\\
	& $L_2$ & \textbf{0.05 (0.01)} & 0.11 (0.00) & 0.11 (0.00)    & 0.44 (0.00)    & 1.12 (0.04) & 0.07 (0.01)          \\
	& $L_3$ & 0.18 (0.01)          & 0.32 (0.01) & 0.31 (0.01) & 0.65 (0.00)    & 1.04 (0.02) & \textbf{0.14 (0.01)} \\
	& $L_4$ & \textbf{0.26 (0.02)} & 0.79 (0.02) & 0.76 (0.02) & 1.31 (0.03) & 2.49 (0.09) & 0.37 (0.06)          \\
	& $L_5$ & \textbf{1.24 (0.04)} & 1.77 (0.04) & 1.75 (0.04) & 1.95 (0.03) & 3.25 (0.17) & 1.44 (0.13)          \\ 	\hline
	p=100 & $L_1$& \textbf{0.06 (0.00)}    & 0.18 (0.01)    & 0.17 (0.02)    & 0.29 (0.01) & 0.85 (0.02) & 0.13 (0.01) \\
	& $L_2$ &\textbf{0.05 (0.01)} & 0.13 (0.01) & 0.12 (0.01) & 0.44 (0.01) & 1.10 (0.02)  & 0.08 (0.01) \\
	& $L_3$ &\textbf{0.18 (0.01)} & 0.34 (0.01) & 0.33 (0.01) & 0.66 (0.02) & 1.04 (0.01) & 0.21 (0.01) \\
	& $L_4$ & \textbf{0.27 (0.02)} & 0.87 (0.01) & 0.82 (0.01) & 1.33 (0.01)& 2.39 (0.05) & 0.59 (0.07) \\
	& $L_5$ & \textbf{1.67 (0.05)} & 2.13 (0.03) & 2.12 (0.03) & 1.97 (0.01) & 3.38 (0.06) & 2.19 (0.15) \\ 	\hline
	p=150 & $L_1$& \textbf{0.07 (0.00)}    & 0.18 (0.01)    & 0.18 (0.01)    & 0.30 (0.00) & 0.86 (0.02) & 0.17 (0.01) \\
	& $L_2$ & \textbf{0.06 (0.01)} & 0.14 (0.01) & 0.13 (0.01) & 0.44 (0.00) & 1.10 (0.02) & 0.14 (0.01)\\
	& $L_3$&\textbf{0.20 (0.01)}  & 0.36 (0.01) & 0.35 (0.01) & 0.66 (0.01) & 1.04 (0.01) & 0.23 (0.01)\\
	& $L_4$ & \textbf{0.29 (0.02)} & 0.91 (0.02) & 0.86 (0.02) & 1.32 (0.02) & 2.41 (0.04) & 0.85 (0.06) \\
	& $L_5$ & \textbf{1.81 (0.05)} & 2.23 (0.03) & 2.23 (0.03) & 1.97 (0.02) & 3.54 (0.13) & 2.98 (0.19)\\ 	\hline
	p=200 &$L_1$ & \textbf{0.07 (0.00)}    & 0.17 (0.01)    & 0.17 (0.00)    & 0.30 (0.01)     & 0.83 (0.02)& 0.22 (0.01)          \\
	&$L_2$ & \textbf{0.08 (0.01)} & 0.15 (0.01) & 0.14 (0.01) & 0.44 (0.00)    & 1.07 (0.02)& 0.16 (0.01)             \\
	&$L_3$ & 0.24 (0.01)          & 0.37 (0.01) & 0.36 (0.01) & 0.66 (0.01)    & 1.02 (0.02) & \textbf{0.18 (0.01)} \\
	&$L_4$ & \textbf{0.33 (0.02)} & 0.92 (0.02) & 0.86 (0.02) & 1.33 (0.02) &2.39 (0.06) & 1.08 (0.08)          \\
	&$L_5$ & \textbf{1.82 (0.03)} & 2.21 (0.02) & 2.21 (0.02) & 1.98 (0.02) & 3.72 (0.07)& 3.72 (0.17)       \\ \hline
	\end{tabular}
	\caption{The loss averages and standard errors (in parenthesis) of estimates when $\Omega_0$ is autoregressive.} \label{table:AR}
\end{table}

\begin{table} 
	\begin{tabular}{cccccccc}\hline
		\multicolumn{1}{l}{} & \multicolumn{1}{l}{} & DAGW.BIC                          & DAGW                              & MLE        & BAYES      & MCD.BIC    & PC-DAG      \\ \hline
	p=30     & $L_1$ & \textbf{0.01 (0.00)}    & 0.01 (0.01)    & 0.02 (0.00)    & 0.02 (0.00)    & 0.25 (0.01) & 0.10 (0.01)           \\
	&$L_2$ & \textbf{0.01 (0.00)}    & 0.01 (0.00)    & 0.01 (0.00)    & 0.01 (0.00)    & 0.25 (0.01) & 0.01 (0.00)             \\
	&$L_3$ & 0.06 (0.01)             & 0.06 (0.02)    & 0.06 (0.02)    & 0.06 (0.01)    & 0.25 (0.01) & \textbf{0.05 (0.01)} \\
	&$L_4$ & \textbf{0.03 (0.00)}    & 0.03 (0.01)    & 0.03 (0.00)    & 0.05 (0.02)    & 0.55 (0.02) & 0.27 (0.04)          \\
	&$L_5$ & \textbf{0.19 (0.01)} & \textbf{0.18 (0.01)} & 0.19 (0.01) & 0.24 (0.01) & 0.88 (0.07) & 1.27 (0.11)        \\ \hline
	p=50     & $L_1$ & \textbf{0.02 (0.00)}    & 0.03 (0.02)    & 0.03 (0.00)   & 0.03 (0.00)    & 0.81 (0.02) & 0.15 (0.01) \\
	&$L_2$ & \textbf{0.02 (0.00)}    & 0.02 (0.01)    & 0.02 (0.00)   & 0.02 (0.00)    & 0.79 (0.02) & 0.02 (0.00)    \\
	&$L_3$ & \textbf{0.11 (0.00)}    & 0.11 (0.01)    & 0.11 (0.00)   & 0.11 (0.01)     & 0.67 (0.01) & 0.12 (0.01) \\
	&$L_4$ & \textbf{0.05 (0.00)}    & 0.05 (0.01)    & 0.05 (0.01)  & 0.07 (0.00)    & 1.71 (0.03) & 0.43 (0.05) \\
	&$L_5$ & \textbf{0.29 (0.01)} & 0.29 (0.03) & 0.30 (0.01) & 0.35 (0.01) & 2.02 (0.07) & 1.83 (0.11)\\ \hline
	p=100    & $L_1$ & \textbf{0.05 (0.01)}    & 0.05 (0.01)   & 0.05 (0.01)  & 0.06 (0.00)    & 1.61 (0.03) & 0.21 (0.01)       \\
	&$L_2$ & 0.05 (0.01)             & 0.05 (0.00)   & 0.05 (0.01) & 0.05 (0.01)    & 1.27 (0.02) & \textbf{0.04 (0.00)} \\
	&$L_3$ & \textbf{0.23 (0.00)}    & 0.26 (0.01)   & 0.26 (0.00) & 0.26 (0.01)   & 1.41 (0.01) & 0.23 (0.01)       \\
	&$L_4$ & \textbf{0.10 (0.00)}     & 0.11 (0.00)    & 0.11 (0.00) & 0.13 (0.01)     & 2.83 (0.03) & 0.65 (0.05)       \\
	&$L_5$ & {0.61 (0.01)} & \textbf{0.6 (0.01)} & 0.61 (0.00) & 0.67 (0.01) & 4.13 (0.09) & 2.68 (0.10)        \\ \hline
	p=150     & $L_1$ & \textbf{0.06 (0.00)} & 0.07 (0.01)             & 0.07 (0.00)    & 0.07 (0.00) & 3.46 (0.09) & 0.26 (0.01) \\
	&$L_2$ & \textbf{0.06 (0.00)} & 0.06 (0.01)             & 0.07 (0.01)    & 0.07 (0.00) & 2.30 (0.02) & 0.06 (0.00)    \\
	&$L_3$ & \textbf{0.39 (0.00)} & 0.39 (0.01)              & 0.39 (0.03)     & 0.39 (0.01)  & 2.57 (0.01) & 0.39 (0.01) \\
	&$L_4$ & 0.15 (0.01)          & \textbf{0.14 (0.00)}    & 0.15 (0.00)    & 0.17 (0.01) & 5.37 (0.02) & 0.80 (0.06)  \\
	&$L_5$ & 0.88 (0.01)          & \textbf{0.87 (0.01)} & 0.88 (0.01) & 0.94 (0.00) & 7.69 (0.12) & 3.30 (0.15)  \\ \hline
	p=200    & $L_1$ & \textbf{0.10 (0.00)}  & 0.12 (0.00)             & 0.12 (0.01)  & 0.11 (0.00)     & 5.16 (0.12)  & 0.29 (0.01) \\
	&$L_2$ & \textbf{0.10 (0.00)}  & 0.10 (0.00)              & 0.11 (0.00)  & 0.11 (0.00)     & 3.06 (0.02)  & 0.12 (0.01)  \\
	&$L_3$ & \textbf{0.62 (0.00)} & 0.62 (0.01)             & 0.63 (0.00) & 0.63 (0.00)    & 3.73 (0.01)  & 0.62 (0.02) \\
	&$L_4$ & \textbf{0.23 (0.00)} & 0.23 (0.00)             & 0.23 (0.00) & 0.26 (0.01)    & 7.38 (0.04)  & 0.86 (0.04) \\
	&$L_5$ & 1.39 (0.01)          & \textbf{1.36 (0.01)} & 1.37 (0.00) & 1.45 (0.01) & 10.75 (0.08) & 3.79 (0.10) \\ \hline 
	\end{tabular}
	\caption{The loss averages and standard errors (in parenthesis) of estimates when $\Omega_0$ has $3\%$ sparsity.} \label{table:Sparse}
\end{table}

\begin{table} 
	\begin{tabular}{cccccccc}\hline
		\multicolumn{1}{l}{} & \multicolumn{1}{l}{} & DAGW.BIC                          & DAGW                              & MLE        & BAYES      & MCD.BIC    & PC-DAG      \\ \hline
		p=30     &$L_1$ & \textbf{0.06 (0.01)} & 0.20 (0.01)           & 0.16 (0.02) & 0.08 (0.01) & 0.21 (0.03) & 0.13 (0.01) \\
		&$L_2$ & 0.34 (0.01)          & \textbf{0.32 (0.02)} & 0.34 (0.04) & 0.33 (0.01) & 0.50 (0.02)  & 0.32 (0.01) \\
		&$L_3$ & 0.72 (0.01)          & \textbf{0.69 (0.02)} & 0.71 (0.06) & 0.70 (0.01)  & 0.86 (0.02) & 0.70 (0.02) \\
		&$L_4$ & 0.71 (0.02)          & \textbf{0.68 (0.03)} & 0.85 (0.08) & 0.74 (0.03) & 1.04 (0.04) & 0.81 (0.03) \\
		&$L_5$ & 1.54 (0.02)          & \textbf{1.52 (0.03)} & 1.78 (0.12) & 1.66 (0.06) & 2.19 (0.04) & 2.17 (0.05)       \\ \hline
		p=50     & $L_1$ & \textbf{0.01 (0.00)} & 0.04 (0.00)    & 0.04 (0.00)    & 0.02 (0.00)    & 0.03 (0.00)    & 0.25 (0.01) \\
		&$L_2$ & \textbf{0.05 (0.00)} & 0.06 (0.00)    & 0.05 (0.00)    & 0.05 (0.00)    & 0.06 (0.00)    & 0.05 (0.00)    \\
		&$L_3$ & \textbf{0.11 (0.00)} & 0.11 (0.00)    & 0.11 (0.00)    & 0.11 (0.00)    & 0.11 (0.00)    & 0.11 (0.00)    \\
		&$L_4$ & \textbf{0.12 (0.00)} & 0.12 (0.00)    & 0.12 (0.00)    & 0.15 (0.01) & 0.16 (0.01) & 0.90 (0.05)  \\
		&$L_5$ & \textbf{0.30 (0.00)}  & 0.31 (0.01) & 0.44 (0.02) & 0.38 (0.02) & 0.87 (0.10)  & 2.89 (0.10) \\ \hline
		p=100    & $L_1$ & \textbf{0.02 (0.00)}    & 0.07 (0.00)    & 0.07 (0.00)             & 0.04 (0.00)    & 0.04 (0.01) & 0.18 (0.01) \\
		&$L_2$ & 0.11 (0.00)             & 0.11 (0.00)    & \textbf{0.09 (0.01)} & 0.10 (0.00)     & 0.12 (0.00)    & 0.10 (0.00)     \\
		&$L_3$ & \textbf{0.22 (0.00)}    & 0.22 (0.00)    & 0.23 (0.01)          & 0.22 (0.00)    & 0.23 (0.00)    & 0.24 (0.00)    \\
		&$L_4$ & \textbf{0.23 (0.00)}    & 0.23 (0.01) & 0.23 (0.02)          & 0.26 (0.01) & 0.28 (0.01) & 0.70 (0.03)  \\
		&$L_5$ & \textbf{0.53 (0.01)} & 0.55 (0.01) & 0.71 (0.04)          & 0.63 (0.02) & 0.94 (0.09) & 2.29 (0.07)      \\ \hline
		p=150     & $L_1$ & \textbf{0.02 (0.00)}    & 0.05 (0.00)    & 0.05 (0.00)    & 0.03 (0.00)    & 0.03 (0.01) & 0.21 (0.01) \\
		&$L_2$ & \textbf{0.07 (0.00)}    & 0.07 (0.00)    & 0.07 (0.00)    & 0.07 (0.00)    & 0.08 (0.00)    & 0.07 (0.00)    \\
		&$L_3$ & \textbf{0.14 (0.00)}    & 0.15 (0.00)    & 0.14 (0.00)    & 0.15 (0.00)    & 0.15 (0.00)    & 0.14 (0.00)    \\
		&$L_4$ & \textbf{0.16 (0.00)}    & 0.16 (0.00)    & 0.16 (0.01) & 0.18 (0.01) & 0.20 (0.01)  & 0.77 (0.03) \\
		&$L_5$ & \textbf{0.38 (0.01)} & 0.39 (0.01) & 0.57 (0.02) & 0.46 (0.02) & 0.88 (0.09) & 2.56 (0.08)  \\ \hline
		p=200    & $L_1$ & \textbf{0.01 (0.00)} & 0.04 (0.00)    & 0.04 (0.00)    & 0.02 (0.00)    & 0.03 (0.00)    & 0.25 (0.01) \\
		&$L_2$ & \textbf{0.05 (0.00)} & 0.06 (0.00)    & 0.05 (0.00)    & 0.05 (0.00)    & 0.06 (0.00)    & 0.05 (0.00)    \\
		&$L_3$ & \textbf{0.11 (0.00)} & 0.11 (0.00)    & 0.11 (0.00)    & 0.11 (0.00)    & 0.11 (0.00)    & 0.11 (0.00)    \\
		&$L_4$ & \textbf{0.12 (0.00)} & 0.12 (0.00)    & 0.12 (0.00)    & 0.15 (0.01) & 0.16 (0.01) & 0.90 (0.05)  \\
		&$L_5$ & \textbf{0.30 (0.00)}  & 0.31 (0.01) & 0.44 (0.02) & 0.38 (0.02) & 0.87 (0.10)  & 2.89 (0.10)  \\ \hline 
	\end{tabular}
	\caption{The loss averages and standard errors (in parenthesis) of estimates when $\Omega_0$ has a compound structure.} \label{table:diag}
\end{table}

\begin{table} 
	\begin{tabular}{cccccccc}\hline
		\multicolumn{1}{l}{} & \multicolumn{1}{l}{} & DAGW.BIC                          & DAGW                              & MLE        & BAYES      & MCD.BIC    & PC-DAG      \\ \hline
		p=30     &$L_1$ & \textbf{0.08 (0.00)} & 0.21 (0.01)          & 0.24 (0.01) & 0.10 (0.00)    & 0.23 (0.03) & 0.14 (0.01) \\
		&$L_2$ & 0.33 (0.01)       & \textbf{0.3 (0.02)}  & 0.38 (0.00)    & 0.37 (0.00)    & 0.50 (0.02)  & 0.30 (0.01)  \\
		&$L_3$ & 0.69 (0.01)       & \textbf{0.65 (0.02)} & 0.75 (0.00)    & 0.75 (0.00)    & 0.85 (0.02) & 0.67 (0.02) \\
		&$L_4$ & 0.67 (0.02)       & \textbf{0.66 (0.03)} & 0.81 (0.01)  & 0.83 (0.00)    & 1.04 (0.04) & 0.83 (0.05) \\
		&$L_5$ & 1.52 (0.04)       & \textbf{1.51 (0.05)} & 1.65 (0.01) & 1.71 (0.01) & 2.22 (0.08) & 2.38 (0.15)     \\ \hline
		p=50     & $L_1$ & \textbf{0.05 (0.00)}    & 0.13 (0.00)             & 0.14 (0.00)    & 0.04 (0.00)    & 0.10 (0.02)  & 0.17 (0.01) \\
		&$L_2$ & 0.20 (0.01)           & \textbf{0.18 (0.01)} & 0.22 (0.00)    & 0.22 (0.00)    & 0.27 (0.01) & 0.19 (0.01) \\
		&$L_3$ & 0.42 (0.01)          & \textbf{0.39 (0.02)} & 0.45 (0.00)    & 0.45 (0.00)    & 0.49 (0.01) & 0.41 (0.01) \\
		&$L_4$ & 0.42 (0.02)          & \textbf{0.41 (0.02)} & 0.48 (0.00)    & 0.50 (0.00)     & 0.58 (0.02) & 0.75 (0.05) \\
		&$L_5$ & \textbf{0.98 (0.03)} & 0.98 (0.04)          & 1.01 (0.01) & 1.07 (0.01) & 1.49 (0.09) & 2.39 (0.12) \\ \hline
		p=100    & $L_1$ & \textbf{0.03 (0.00)}    & 0.07 (0.00)            & 0.07 (0.00)    & 0.05 (0.00)    & 0.05 (0.01) & 0.23 (0.01) \\
		&$L_2$ & \textbf{0.10 (0.00)}     & 0.10 (0.00)             & 0.11 (0.00)    & 0.11 (0.00)    & 0.12 (0.00)    & 0.10 (0.00)     \\
		&$L_3$ & 0.21 (0.00)             & \textbf{0.20 (0.01)} & 0.22 (0.00)    & 0.22 (0.00)    & 0.24 (0.00)    & 0.21 (0.00)    \\
		&$L_4$ & \textbf{0.22 (0.01)} & 0.22 (0.01)         & 0.24 (0.00)    & 0.26 (0.00)    & 0.29 (0.02) & 0.85 (0.06) \\
		&$L_5$ & \textbf{0.54 (0.02)} & 0.56 (0.02)         & 0.57 (0.01) & 0.59 (0.01) & 1.03 (0.11) & 2.89 (0.12)   \\ \hline
		p=150     &$L_1$ & \textbf{0.02 (0.00)}    & 0.05 (0.00)          & 0.05 (0.00) & 0.02 (0.00)    & 0.04 (0.00)    & 0.26 (0.01) \\
		&$L_2$ & 0.07 (0.00)             & \textbf{0.06 (0.00)} & 0.08 (0.00) & 0.08 (0.00)    & 0.08 (0.00)    & 0.07 (0.00)    \\
		&$L_3$ & \textbf{0.14 (0.00)}    & 0.14 (0.00)          & 0.15 (0.00) & 0.15 (0.00)    & 0.15 (0.00)    & 0.14 (0.00)    \\
		&$L_4$ & \textbf{0.15 (0.01)} & 0.15 (0.01)       & 0.16 (0.00) & 0.18 (0.00)    & 0.21 (0.01) & 0.98 (0.04) \\
		&$L_5$ & \textbf{0.39 (0.01)} & 0.40 (0.02)        & 0.38 (0.00) & 0.43 (0.01) & 0.92 (0.12) & 3.34 (0.09)\\ \hline
		p=200  &$L_1$ & \textbf{0.01 (0.00)} & 0.04 (0.00) & 0.04 (0.00) & 0.02 (0.00)    & 0.03 (0.00)    & 0.30 (0.01)  \\
		&$L_2$ & \textbf{0.05 (0.00)} & 0.05 (0.00) & 0.06 (0.00) & 0.06 (0.00)    & 0.06 (0.00)    & 0.05 (0.00)    \\
		&$L_3$ & \textbf{0.11 (0.00)} & 0.11 (0.00) & 0.11 (0.00) & 0.11 (0.00)    & 0.11 (0.00)    & 0.11 (0.00)    \\
		&$L_4$ & \textbf{0.12 (0.00)} & 0.12 (0.00) & 0.12 (0.00) & 0.14 (0.00)    & 0.17 (0.01) & 1.14 (0.07) \\
		&$L_5$ & \textbf{0.32 (0.00)} & 0.33 (0.00) & 0.33 (0.00) & 0.35 (0.01) & 0.99 (0.06) & 3.72 (0.14)\\ \hline 
	\end{tabular}
	\caption{The loss averages and standard errors (in parenthesis) of estimates when $\Omega_0$ has a permuted compound structure.} \label{table:diag_perm}
\end{table}

	\section{Posterior convergence rate for DAG-Wishart priors} \label{sec:posterior}
		In this section, we will provide the convergence rate for the posterior distribution of the 
	precision matrix under the DAG-Wishart prior. 	We 
	assume that the data matrix $\bm Y$ is actually being generated from a true model obeying $\mbox{MN}_{n\times p}\left(0,    I_n,    \Sigma_0^n\right)$, where $\Sigma_0^n = (\Omega_0^n)^{-1}=L_0^n(D_0^n)^{-1}(L_0^n)^T$. Denote $d_n$ as the maximum number of 
	non-zero entries in any column of the true Cholesky factor $L_0^n$. In order to establish our asymptotic results, we need the following mild regularity 
	assumptions. Each assumption below is followed by an interpretation/discussion. 
	Note that for a symmetric $p \times p$ matrix $A = (A_{ij})_{1\le i,j\le p}$, let $eig_1(A) 
	\le eig_2(A) \ldots eig_p(A)$ denote the ordered eigenvalues of $A$. 
	\begin{assumption}
		There exists $\epsilon_{0} \le 1$, such that for every $n \ge 1,$ $0 < \epsilon_{0} \le eig_1({\Omega}_0^n) \le eig_{q_n}({\Omega}_0^n) \le \epsilon_{0}^{-1}$.
	\end{assumption}
	\noindent
	This assumption ensures that the eigenvalues of the true precision matrices are bounded by fixed constants, which has been commonly used for establish high dimensional covariance asymptotic properties. See for example \citep{Bickel:Levina:2008, ElKaroui:2008, Banerjee:Ghosal:2014, XKG:2015, Banerjee:Ghosal:2015}. \citet{CKG:2017} relax this assumption by allowing the lower and upper bounds on the eigenvalues to depend on $p$ and $n$. 
	\begin{assumption}
		$d_n^{2} \sqrt{\frac{\log p_n}{n}}\rightarrow 0$, as $n \rightarrow \infty$. 
	\end{assumption}
	\noindent
	This assumption essentially states that the number of variables $p_n$ has to 
	grow slower than $e^{n/d_n^{4+2k}}$ (and also $e^{n/(\log n)^{2+k}}$). Again, 
	similar assumptions are common in high dimensional covariance asymptotics, 
	see for example 
	\cite{Bickel:Levina:2008, XKG:2015, Banerjee:Ghosal:2014, Banerjee:Ghosal:2015}. 
	\begin{assumption}
		For every $n \ge 1$, the hyperparameters for the DAG-Wishart prior 
		$\pi_{U_n,\bm{\alpha}( \mathcal{D}_n)}^{\Theta_{ \mathcal{D}_n}}$ in (\ref{a3}) satisfy 
		(i) $2 < \alpha_i( \mathcal{D}_n) - \nu_i( \mathcal D_n) < c$ for every $ \mathcal{D}_n$ 
		and $1 \le i \le p_n$, and (ii) $0 < \delta_1 \le eig_1(U_n) \le eig_{p_n}(U_n) \le 
		\delta_2 < \infty$. Here $c, \delta_1, \delta_2$ are constants not depending on $n$. 
	\end{assumption}
	
	\noindent
	This assumption provides mild restrictions on the hyperparameters for the 
	DAG-Wishart distribution. The assumption $2 < \alpha_i( \mathcal D) - 
	\nu_i( \mathcal D)$ establishes prior propriety. The assumption $\alpha_i( \mathcal D) - 
	\nu_i( \mathcal D) < c$ implies that the shape parameter $\alpha_i( \mathcal D)$ can only differ 
	from $\nu_i( \mathcal D)$ (number of parents of $i$ in $ \mathcal D$) by a constant which 
	does not vary with $n$. Additionally, the eigenvalues of the 
	scale matrix $U_n$ are assumed to be uniformly bounded in $n$. 
	
Let $|| A ||_{(2,2)} = 
	\{eig_p(A^TA)\}^{\frac12}$, $||A||_F = \{\mbox{tr}(AA^T)\}^{\frac 1 2}$ represent the 2-norm and Frobenius norm respectively for any $p \times p$ matrix $A$, and 
	$\Pi(\cdot \; \mid \bm{Y})$ denote the probability measure corresponding to the 
	posterior distribution. Let $\bar P$ and $\bar{E}$ respectively denote the probability 
	measure and expected value corresponding to the ``true" Gaussian DAG model. For sequences  $a_n$ and $b_n$, $a_n \sim b_n$ means $\frac{a_n}{b_n} \rightarrow c$ for some constant $c > 0$. We now present our two results on the posterior convergence rates for both the precision matrix and the Cholesky factors under a given permutation and multiple ensemble estimates in Theorem \ref{3.1} and \ref{3.2}, similar to that in \citep*{CKG:2017}.
	\begin{theorem} \label{3.1}
		Let $\Omega_\sigma = L_\sigma D_\sigma^{-1} L_\sigma^T$ be the posterior under the DAG-Wishart distribution with respect to a variable order $\sigma$. Under Assumption 1-3 above, for a large enough constant $K$(not depending on $n$), the posterior distributions for $\Omega_\sigma$, $L_\sigma$ and $D_\sigma$ satisfies:
		$$
		\bar{E} \left [\Pi \left\{|| {\Omega_\sigma} - \Omega_{0_\sigma} ||_{(2,2)} \ge K d^{2}\sqrt{\frac{\log{p}}{n}} | \bm{Y} 
		\right\} \right] \rightarrow 0,
		$$
		
		\begin{equation*}
			\bar{E} \left [\Pi \left\{|| { L_\sigma} -  L_{0\sigma} ||_{(2,2)} \ge K d^{2}\sqrt{\frac{\log{p}}{n}} | \bm{Y} 
		\right\} \right] \rightarrow 0,
		\end{equation*}
		
		\noindent
		and 
		\begin{equation*}
	\bar{E} \left[\Pi \left\{|| D_\sigma^{-1}-{D_{0\sigma}}^{-1}|| _{(2,2)} \ge K d^{2}\sqrt{\frac{\log{p}}{n}} |\bm{Y}\right\} \right] {\rightarrow} 0,
		\end{equation*}
		as $n \rightarrow \infty$. 
	\end{theorem}
\noindent
 Note that Theorem \ref{3.1} provides the convergence rate for the posteriors under a given permutation. In order to establish the consistency results for our ensemble estimate that incorporates the information from multiple permutations, we present the following result.
\begin{theorem} \label{3.2}
Under Assumption 1-3, if we further assume the hard thresholding parameter adopted in Step 5 in Algorithm \ref{final_algorithm} $\tau_b$ satisfies $\tau_b  \sim \sqrt{\frac{\log p} {nM}}$, then the final estimate $\check{\Omega}_{\tau_b}$ satisfies
	$$
\bar{E} \left [\Pi \left\{|| \check{\Omega}_{\tau_b} - \Omega_{0} ||_{F} \ge K d^{2}\sqrt{\frac{p^2\log{p}}{nM}} | \bm{Y} 
\right\} \right] \rightarrow 0,
$$
where M is the number of permutations generated for obtaining the ensemble estimate. 
\end{theorem}
Theorem \ref{3.1} and \ref{3.2} immediately follows by noting $||A||_F \le \sqrt{p}||A||_{(2,2)}$ for any $p\times p$ nonsingular matrix.

\section{Discussion} \label{sec:discussion}
In this paper, we propose a novel permutation-based Bayesian approach for estimating the inverse covariance matrix under DAG-Wishart distributions. For each permutation, we first estimate the true underlying DAG using graph selection procedures proposed by~\citet*{CKG:2017}. Then based on each estimated DAG, we obtain the MAP estimate for the Cholesky factor under DAG-Wishart distribution. The final estimate is constructed by taking average over all the estimates from permutations. Further thresholding procedures can be implemented to promote sparsity. It is worthwhile to point out that the proposed estimator does not require the ordering of variables to be known and therefore, could serve as a more flexible, yet precise estimate for the inverse covariance matrix, as indicated by the simulation studies.

\section{Acknowledgments}
We would also like to thank the reviewers for their helpful and constructive comments which substantially improve the quality of the paper.
\bibliographystyle{plainnat}
\bibliography{references}
	
\end{document}